 \documentclass[final,5p,times,twocolumn,authoryear]{elsarticle}

\usepackage{amssymb}
\usepackage{multirow}

\journal{Planetary and Space Science}

\begin{document}

\begin{frontmatter}



\title{Extrasolar comets : the origin of dust in exozodiacal disks ?}


\author{U.~Marboeuf}
\ead{ulysse.marboeuf@space.unibe.ch}
\address{Physics Institute, University of Bern, Bern, Switzerland}
\address{Univ. Grenoble Alpes, IPAG, F-38000 Grenoble, France}
\address{CNRS, IPAG, F-38000 Grenoble, France}

\author{A.~Bonsor}
\address{Institute of Astronomy, University of Cambridge, Madingley Road, Cambridge CB3 0PA, United Kingdom}
\address{Univ. Grenoble Alpes, IPAG, F-38000 Grenoble, France}
\address{CNRS, IPAG, F-38000 Grenoble, France}

\author{J.-C.~Augereau}
\address{Univ. Grenoble Alpes, IPAG, F-38000 Grenoble, France}
\address{CNRS, IPAG, F-38000 Grenoble, France}



\begin{abstract}
Comets have been invoked in numerous studies as a potentially important source of dust and gas around stars, but none has studied the thermo-physical evolution, out-gassing rate, and dust ejection of these objects in such stellar systems.
In this paper we investigate the thermo-physical evolution of comets in exo-planetary systems in order to provide valuable theoretical data required to interpret observations of gas and dust.
We use a quasi 3D model of cometary nucleus to study the thermo-physical evolution of comets evolving around a single star from 0.1 to 50 AU, whose homogeneous luminosity varies from 0.1 to 70 L$_\odot$.
This paper provides thermal evolution, physical alteration, mass ejection, lifetimes, and the rate of dust and water gas mass productions for comets as a function of the distance to the star and stellar luminosity. 
Results show significant physical changes to comets at high stellar luminosities.
The mass loss per revolution and the lifetime of comets depend on their initial size, orbital parameters and follow a power law with stellar luminosity. The models are presented in such a manner that they can be readily applied to any planetary system. 
By considering the examples of the Solar System, Vega and HD 69830, we show that dust grains released from sublimating comets have the potential to create the observed (exo)zodiacal emission. We show that observations can be reproduced by 1 to 2 massive comets or by a large number of comets whose orbits approach close to the star.
Our conclusions depend on the stellar luminosity and the uncertain lifetime of the dust grains. We find, as in previous studies, that exozodiacal dust disks can only survive if replenished by a population of typically sized comets renewed from a large and cold reservoir of cometary bodies beyond the water ice line. These comets could reach the inner regions of the planetary system following scattering by a (giant) planet. 
\end{abstract}


\end{frontmatter}


\section{Introduction}

In the solar system, comets are thought to be the most primitive bodies. These small porous objects, composed of a mixture of different ices (H$_2$O, CO, CO$_2$, ...) and refractory elements, were formed in cold areas of the protoplanetary disk. 
A specific peculiarity to comets is their eccentric orbits due to single or mutiple scattering events by planets \footnote{The existence of eccentric orbits need gravitational perturbation: Jupiter Family Comets in the solar system exist due to multiplanet captures from the transneptunien population (see Jewitt 2004; Duncan et al. 2004).}. During each perihelion passage, whilst the comets are close to the Sun, they undergo thermo-physical changes due to the sublimation of ices, and consequently the ejection of gaseous volatile species and dust grains from their surfaces.
In the inner solar system, the splitting of comets accounts for 85\% of the zodiacal dust (Nesvorny et al. 2010).

There is significant evidence for the presence of cometary-type bodies in exo-planetary systems. 
Belts of dust and rocks known as debris disks have been observed around a large fraction of nearby stars (33\% of nearby A stars according to Su et al. 2006, $\sim$20\% of nearby solar type stars according to Eiroa et al. 2013, and at least 15\% of mature stars, 10 Myr to 10 Gyr, according to Moro-Martin 2013). Although it is the thermal emission from small dust particles that is observed, the presence of a population of larger bodies is inferred from the short lifetime of the dust grains ($\leq$1 Myr, see Moro-Martin 2013) and the need to replenish them continuously in a collisional cascade (Wyatt et al. 2007). These debris disks are massive equivalents to the solar system's Kuiper belt. 
There is, however, no good reason to suppose that scattering processes, similar to those that may have scattered Kuiper belt objects inwards to become Jupiter Family Comets (e.g. Levison \& Duncan 1997), do not occur in exo-planetary systems.

Observational evidence for a population of comets on eccentric orbits in exo-planetary systems is difficult to come by, given that only high levels of small dust grains and/or variable gaseous absorption features in debris disks (Montgomery \& Welsh 2012) are detectable. There are, however, several unusual observations of exo-planetary systems that could potentially be related to comets on eccentric orbits. High levels of exozodiacal dust have been detected around a large fraction of nearby Main Sequence stars independently of their age (29\% Absil et al. 2013; Ertel et al. 2014). Such high levels of dust cannot be explained by steady-state collisional evolution (Wyatt et al. 2007) and it may be that we are witnessing evidence of comets on eccentric orbits. 
Another possible signature of such eccentric bodies comes from the pollution and dusty/gaseous circumstellar material observed around many dwarfs (e.g. Farihi et al. 2009; Gaensicke et al. 2006; Zuckerman et al. 2003, 2010; Beichman et al. 2005). The composition of the polluting material resembles planetary material (Klein et al. 2010; Girven et al. 2012) and is thought to originate from comets/asteroids scattered onto star-grazing orbits following stellar mass loss (Debes \& Sigurdsson 2002; Jura 2008; Bonsor et al. 2011, 2012; Debes et al. 2012; Raymond \& Bonsor 2014).

There are also a lot of studies which refer to comets as a source of dust in debris disks around Beta Pictoris (Kifer et al. 2014b; Beust \& Morbidelli 2000; Karmann et al. 2001, 2003; Th\'ebault \& Beust 2001; Beust et al. 2001; Li \& Greenberg 1998; Beust \& Morbidelli 1996; Lecavelier des Etangs et al. 1996; Beust \& Lissauer 1994; Beust et al. 1991, 1990, 1989). Transient phenomena observed on stars have been attributed to infalling exocomets (Alcock et al. 1986; Grady et al. 1996; Roberge et al. 2002; Welsh \& Montgomery 2013) or to orbiting comets around stars (Saavik Ford \& Neufeld 2001; Melnick et al. 2001; Stern et al. 1990). Comets can make themselves known through their long tails of gas and debris that comes off as they approach the stars (de Vries et al. 2012). Effects of photometric variations by cometary tail have already been observed with high resolution spectroscopy (Hainaut 2011) and studied by Lamers et al. (1997), Lecavelier des Etangs et al. (1999a, 1999b), and Jura (2005a, 2005b). The detection of water vapor (Melnick et al. 2001) and OH (Ford et al. 2003) around the star IRC + 10216 has been interpreted by authors as evidence for the existence of an extrasolar cometary system (Ford 2004). Moreover, the presence of a debris disk with a gaseous component around the stars HD172555, HD 21620, HD 110411, HD 145964, and HD 183324 could be due to infalling exocomets (Kiefer et al. 2014a; Welsh \& Montgomery 2013).

In the frame of these studies, a numerical study of the outgassing and dust ejection by exocomets around star of different luminosities seems to be of great interest to give a first estimation of the number of objects needed to account for the dust mass in hot/warm debris disks (exozodiacal disks, or exozodis) and the lifetime of such bodies.
We focus this work on the sublimation of comets in exo-planetary systems made of a single star, with a particular focus on the release of water gas and dust during the perihelion passage, with relevance for its observability in exo-planetary systems. We use a cometary nucleus model initially developed for the thermal evolution of comets in the solar system (Marboeuf et al. 2012; Marboeuf \& Schmitt 2014) and investigate its application to comets in other planetary systems with different stellar luminosities.

This article is organized as follows: Sect.~\ref{sec:models} is devoted to a short presentation of the cometary nucleus model used and the description of the main physical processes taken into account. In Sect.~\ref{paramsection}, we discuss the physical assumptions and thermodynamics parameters adopted for exo-comets. In Sect.~\ref{results} we present results about thermo-physical evolution, and dust and water gas productions of a single comet around stars of different luminosities. Sect.~\ref{disks} is devoted to comparison with debris disks. We finally discuss and summarize our results in Sect.~\ref{discussion}.

\section{The quasi-3D cometary model \label{sec:models}}

\begin{figure}
\includegraphics[width=9.cm]{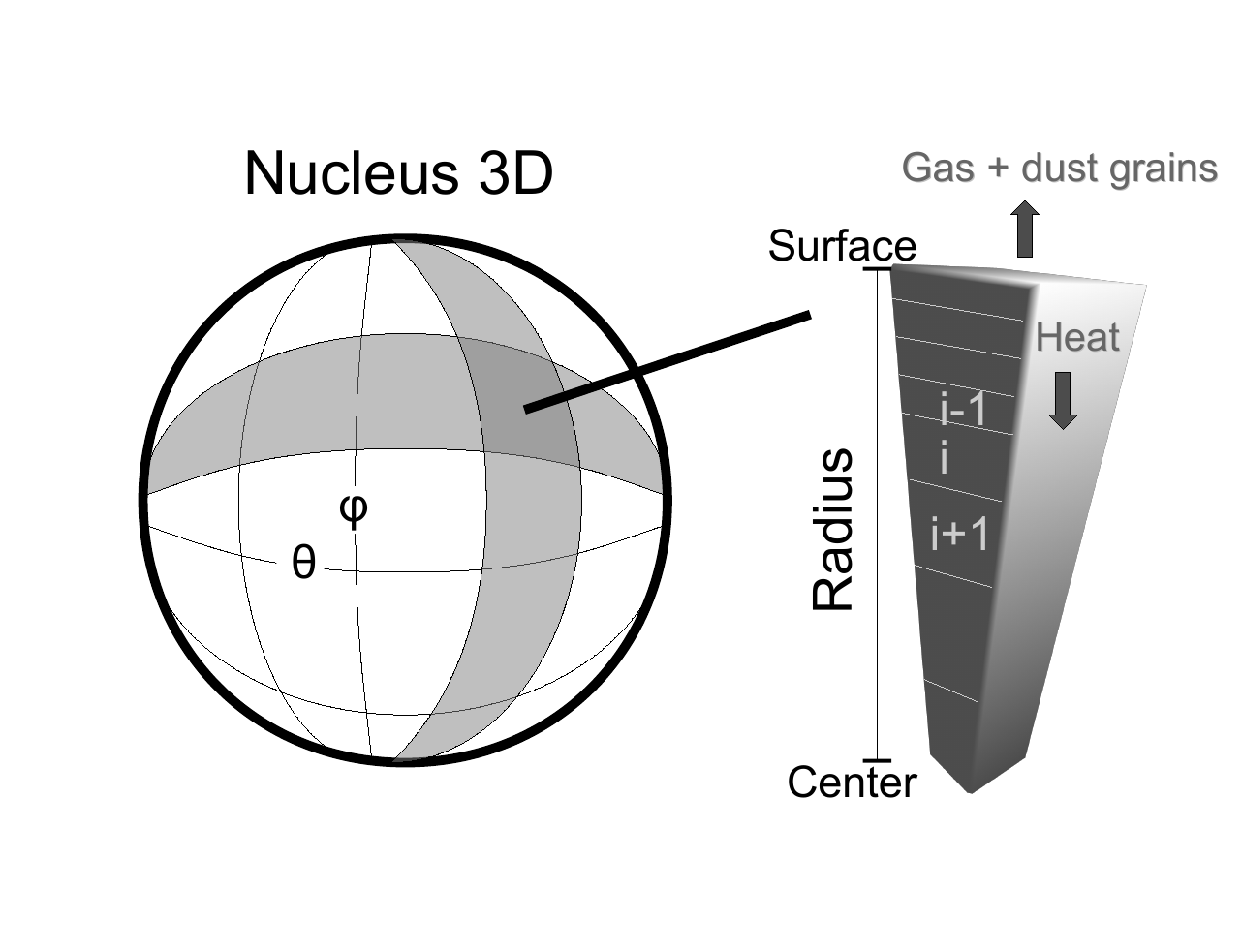}
\caption{Schematic view of the quasi 3D nucleus model of comet. Heat
  conduction and gas diffusion occur only radially throughout the
  nucleus. From Marboeuf \& Schmitt (2014).}
\label{fig:3D}
\end{figure}

The cometary material is believed to be composed of dust grains with an icy mantle formed by water and some other volatile molecules (see Fig.1 in Marboeuf et al. 2012). The interior of comets is modeled by a porous predefined mixture (hereafter porous icy matrix) composed of water ice, volatile species (in gas and solid states), and dust grains (embedded in water ice) in specified proportions. 

The numerical model presented in this work uses the quasi 3D approach (see Marboeuf \& Schmitt 2014). It allows us to take into account spatial variations of the temperature on the surface of comets due to the unevenly distributed solar radiation over the cometary surface (see Rosenberg \& Prialnik 2007).
This type of model is able to handle both the diurnal latitudinal and longitudinal variations of illumination of the surface of the rotating
comet (see e.g. Gutie\'rrez et al. 2000; Julian et al. 2000; Cohen et al. 2003; Lasue et al. 2008; Rosenberg \& Prialnik 2007) during its revolution around the star. This leads to a much better estimation of the temperature on its surface compared to the 1D model (Marboeuf et al. 2012) which considers an average temperature everywhere on the surface of the comet with spherical symmetry whatever the erosion of the nucleus. The quasi 3D is of particular interest when comets approach the star since the temperature varies greatly as a function of the latitude and longitude (day side, night side). Moreover, we note that the quasi 3D approach at low thermal conductivity such as in comets is a good approximation of the fully 3D model regarding timescales (see Prialnik et al. 2004) of heat diffusion through the nucleus (between some months to several years on 1~m thickness material, depending on the thermal inertia of the material), and those on the sublimation of water ice, and insolation (both between a few second and some minutes).
The quasi 3D model represents a spherical comet whose surface is divided numerically in several sections ($\varphi$, $\theta$) as illustrated in
Fig.~\ref{fig:3D}, and below which the interior of the comet is divided in several radial layers ($i$ index) whose thickness follows initially a power law. The size and number of these radial layers can increase or decrease during the lifetime of the comet following its erosion (sublimation of ices and dust grain ejection).
We note that each section ($\varphi$, $\theta$) evolves independently of the others and only radial flow of gas and heat are considered in the model. 
The initial spherical shape assumed in this model does not correspond to shape properties observed for comets of the solar system since the small bodies have negligible self-gravity. However, it is impossible to determine the exact shape of a comet and the spherical approximation of the shape allows us to simplify the calculation at the beginning of the computation. Nevertheless, the independence of the physical change of facets of the comet allows to modify substantially the shape of the comet with time, despite that meridional and azimuthal heat and gas fluxes are not calculated.

The model is able to reproduce the thermo-physical behavior of the cometary material when the comet approaches the star (see Marboeuf \& Schmitt 2014). It can take into account several volatile species together and several water ice structures (amorphous ice or clathrates with trapped gases, pure crystalline, or a mixture of these structures; see Marboeuf et al. 2012).  
However, we consider only H$_2$O ice as volatile species in comets of this study in order to avoid excessively long computing time (see Sect.~\ref{paramsection}).
The initial structure of H$_2$O ice in the model is then chosen as crystalline since no gas species ($\neq$ of H$_2$O) are included in the study, and consequently in the water ice structure. Moreover, as shown in Marboeuf \& Schmitt (2014), the initial structure of water ice does not change the outgassing rate of water.

This quasi 3D model describes radial heat transfers (see Sec.~\ref{NRJ} and Eq.~\ref{NRJ_cons}), latent heat exchanges, 
sublimation/condensation of H$_2$O, and radial gas diffusion through the porous network of the nucleus (see Sec.~\ref{Mass} and Eq.~\ref{MASS_cons}). 
At the surface of the nucleus, the model takes into account the water gas and dust grains ejection. 
We note that dust grains ejection can only occur once they are free. It is the water sublimation which controls the dust grain ejection.
Descriptions and assumptions on physical processes taken into account in the model are fully explained in Marboeuf et al. (2012) and Marboeuf \& Schmitt (2014).
Hereafter, we provide a simple description of main physical processes taken into account in the model.

\subsection{Energy conservation \label{NRJ}}

For each layer $i$ and position ($\varphi$, $\theta$) in the nucleus, we solve the energy conservation equation that describes the radial heat diffusion through the porous matrix:
\begin{equation} 
  \rho c \frac{\partial T}{\partial t}= \nabla. \left(K^{m} \frac{\partial
 T}{\partial r}\right) - \sum_{x} H_x^s Q_{x} +Y   \qquad (J \, m^{-3} \, s^{-1}) 
\label{NRJ_cons}
\end{equation}
with $\rho$ (kg m$^{-3}$) the density of solids, $c$ (J kg$^{-1}$ K$^{-1}$) their specific heat capacity, $T$ (K) the temperature, $t$ (s) the time, and $r$ (m) the distance to the center of the nucleus. $H_x^s$ (J mol$^{-1}$) is the molar latent heat of sublimation of the ice $x$ and $Q_x$ (mol m$^{-3}$ s$^{-1}$) represents the rate of moles of gas $x$ per unit volume that sublimates$/$condenses in the porous network or/and that is released by amorphous ice during the process of crystallization. Its expression is given by the inversion of the gas diffusion equation~(\ref{MASS_cons}) given below.
$Y$ represents the power per unit volume released during the crystallization process of amorphous water ice (see Espinasse et al. 1991), exchanged between the gas phase (which diffuse in the porous network) and the solid matrix, or/and released/taken during the formation/dissociation of cages of clathrate.
$K^{m}$ (J s$^{-1}$ m$^{-1}$ K$^{-1}$) is the heat conduction coefficient of the porous matrix whose expression is given by Hertz's formula (see Kossacki et al. 1999; Davidsson \& Skorov 2002; Prialnik et al. 2004; Huebner et al. 2006; Marboeuf et al. 2012):
\begin{equation}
K^{m}=h K^s  \qquad \textstyle {(W \, m^{-1} K^{-1})}
\label{ctc}      
\end{equation}
where $K^s $ is the conductivity of the solid (dust and ices) components (see Marboeuf et al. 2012). $h$ is the Hertz factor used to correct the effective area of the matrix material through which heat flows (Davidsson \& Skorov 2002; Prialnik et al. 2004). $h$ is expressed by considering two spheres of radius $R$ that are pressed together and have a contact area of radius $r_c$ ($h\approx \frac{r^2_c}{R^2}$, Kossacki et al. 1999). Its value can vary between 10$^{-3}$ and 10$^{-1}$ (see Davidsson \& Skorov 2002; Huebner et al. 2006, Volkov \& Lukyanov 2008).

\subsection{Mass conservation \label{Mass}}
For each layer $i$, position ($\varphi$, $\theta$) and molecule $x$, we solve the diffusion of gas through the porous matrix by using the mass conservation equation:
\begin{equation}
  \frac{\partial \rho_x^g }{\partial t}=M_x \left(\nabla .\left (\Phi_x \right) +Q_x \right) \qquad (kg \, m^{-3} \, s^{-1})
\label{MASS_cons}
\end{equation}
where $Q_x$ (mol m$^{-3}$ s$^{-1}$) represents the net source of gas $x$ released in the porous network during water ice crystallization, the pure ice sublimation/condensation, or/and the rate of clathrate dissociation/formation. $\rho_x^g$ (kg m$^{-3}$) is the mass density of gas $x$, and $M_x$ (kg mol$^{-1}$) its molar mass. $\Phi_x$ (mol m$^{-2}$ s$^{-1}$) is the molar flow through the porous network whose expression is given in Marboeuf et al. (2012).

\subsection{Surface erosion and rate production of water at the surface of comets \label{appendix}}

\subsubsection{Calculation of the temperature and ice sublimation at the surface of the nucleus \label{tempsurface}}

The sublimation of water and ejection of dust grains from the nucleus
is mainly function of the temperature $T_{\varphi, \theta}$ of the
surface element at the position ($\varphi$, $\theta$). Its accurate
calculation is therefore of great importance. The temperature of each
section ($\varphi$, $\theta$) of the cometary surface is given by a
thermal balance between the solar energy absorbed by the cometary
material (on the left part of the equation) and its thermal emission,
the heat diffusion towards the interior of the nucleus and the energy
of sublimation of water ice\footnote{By considering only this species in the nucleus, see Sect.\ref{chemicalcompo}.} (on the right side of the equation) existing on the elemental surface ($\varphi$, $\theta$):
\begin{equation} 
\frac{C_s L_* (1-A_b)}{r_h^2} ~\zeta_{\varphi, \theta} = \epsilon \sigma
T_{\varphi, \theta}^4 + K \frac{\partial T_{\varphi, \theta}}{\partial
  r} + \alpha^{\varphi, \theta}_{H_2O} H_{H_2O}^s \Phi_{H_2O}(T_{\varphi, \theta}) \qquad (W \, m^{-2})
\label{boundary_heat_surf}
\end{equation}
where $C_s$ (W m$^{-2}$) is the solar constant, $L_*$ the stellar luminosity (in $L_\odot$), $A_b$ the Bolometric Bond Albedo, $r_h$ (AU) the distance to the star, $\epsilon$ the infrared surface emissivity, $\sigma$ the Stefan-Boltzmann constant (W m$^{-2}$ K$^{-4}$), $T_{\varphi, \theta}$ (K) the temperature of the section ($\varphi$, $\theta$) of the surface, and $\alpha^{\varphi, \theta}_{H_2O}$ the surface fraction covered by water ice. $\Phi_{H_2O}(T_{\varphi, \theta})$ (mol m$^{-2}$ s$^{-1}$) is the free sublimation rate of water in vacuum given by the expression of Delsemme $\&$ Miller (1971):
\begin{equation}
  \Phi_{H_2O}(T_{\varphi, \theta})=\frac{P_{H_2O}^s(T_{\varphi, \theta})}{\sqrt{2\pi M_{H_2O}RT_{\varphi, \theta}}} \qquad (mol \, m^{-2} \, s^{-1})
  \label{phi}
\end{equation}
where $M_{H_2O}$ (kg mol$^{-1}$) is the molar mass of water gas, $R$ the perfect gas constant (J mol$^{-1}$ K$^{-1}$), and $P_{H_2O}^s(T_{\varphi, \theta})$ the water vapor sublimation pressure (Pa) given in Fray \& Schmitt (2009):
\begin{equation} 
P_{H_2O}^s(T_{\varphi, \theta}) = P_t \times e^{\frac{3}{2} ln(\frac{T}{T_t}) + (1-\frac{T_t}{T}) \times \eta(\frac{T}{T_t})} \qquad (bar)
\label{eq:pressurewater}
\end{equation}
with T$_t=$273.16 K, P$_t=$ (6.116577$\pm$0.0001)$\times$10$^{-3}$ bar, and $\eta(\frac{T}{T_t})$ given by:
\begin{equation} 
\eta(\frac{T}{T_t}) = \sum^{6}_{i=0} e_i (\frac{T}{T_t})^i
\label{eq:polynome}
\end{equation}
The coefficients of the polynomial $\eta(\frac{T}{T_t})$ are given in Tab.~\ref{param_polynom}. We note that Eq.~(\ref{eq:pressurewater}) is only valid for temperatures lower than triple point $T_t$.
\begin{table}
\centering 
\caption{Coefficients of the polynomial $\eta(\frac{T}{T_t})$ (Eq.~\ref{eq:polynome}) for the sublimation pressure equation (\ref{eq:pressurewater}) of water ice.}
\begin{tabular}{|l|c|}
\hline
i & e$_i$ 			  \\ 
\hline
   0	& 20.9969665107897 \\
   1  & 3.72437478271362  \\
   2  & -13.9205483215524  \\
    3  & 29.6988765013566  \\
     4  & -40.1972392635944  \\
      5  & 29.7880481050215  \\
       6  & -9.13050963547721  \\
  \hline
\end{tabular}
\label{param_polynom}
\end{table}

$\zeta_{\varphi, \theta}$ is the term of insolation of a facet at the
position ($\varphi$, $\theta$) of the surface of the nucleus.  We use
the "slow-rotator" approach that takes into account the diurnal
latitudinal and longitudinal variations of illumination on the facets
of the surface. This approach allows to obtain an accurate surface
temperature distribution and its diurnal changes at any heliocentric
distance (Prialnik 2004).  With this approach, $\zeta_{\varphi,
  \theta}$ is equal at $max(\cos~\xi, 0)$ in the
Eq.~(\ref{boundary_heat_surf}), with $\xi$ the stellar zenith distance
calculated as (see Sekanina 1979; Fanale \& Salvail 1984; Prialnik
2004; Gortsas et al. 2011):
\begin{equation}
\cos~\xi = \cos~\theta ~\cos\left(\omega~(t-t_0) \right)~\cos~\theta_s + \sin~\theta ~\sin~\theta_s
\end{equation}
with $\theta$ the latitude on the comet, $t$ (s) the time since the
beginning of the computation, $t_0$ (s) the initial time of
computation, $\theta_s$ the cometocentric latitude of the sub-solar
point that takes into account the obliquity of the comet (see Marboeuf
et al. 2012), and $\omega=\frac{2\pi} {P_r}$ where $P_r$ (s) is the
nucleus rotation period of the comet.

\subsubsection{Surface erosion and dust grains ejection at the surface of the nucleus \label{dustejection}}

At the beginning of the computation, the grains are encased in water ice and the comet has a homogeneous physical composition. 
The size distribution of dust grains encased in H$_2$O ice is given by a power law (Rickman et al. 1990):
\begin{equation}
N(a) {\rm d}a= N_0 a^{\beta} {\rm d}a
\end{equation}
where $\beta$ is the power law index of the size distribution and $N_0$ a normalization factor.
By approaching the star, the temperature of surface increases. H$_2$O ice begins then to sublimate and grains can then be freed. At this time, the variation of the radius $\Delta R_n^{\phi, \theta}$ of the nucleus at the latitude $\theta$ and longitude $\varphi$ is then recomputed by
using the following equation:
\begin{equation}
\Delta R_n^{\phi, \theta} = \sum \frac{M_{H_2O} \Phi_{H_2O}(T_{\varphi, \theta})}{\rho^{\varphi, \theta}} \Delta t \qquad (m)
\label{erosion_comet_water}
\end{equation}
where $\Phi_{H_2O}^{\varphi, \theta}$ (kg m$^{-2}$ s$^{-1}$) and $\rho^{\varphi, \theta}$ (kg m$^{-3}$) are respectively the flow of water gas (see Eq.~\ref{phi} in Sect.~\ref{tempsurface}) and mass density of solid at the position ($\varphi$, $\theta$) of the surface of the comet, and $\Delta t$ the time step of the computation (see Eq.~\ref{deltat} in Sect.~\ref{timestep2}). 
In this model, all dust grains freed by the sublimation of water ice are fully ejected in space from the surface of each facet ($\varphi$, $\theta$) of the nucleus. \\

We note that the size and the number of cells are recalculated at each time step after erosion in order to keep small thicknesses of layers near the surface of each section ($\varphi$, $\theta$) of the nucleus following the method described in Marboeuf et al. (2012).

\subsubsection{Orbital position and time step \label{timestep2}}
At the end of each time step, the orbital position $r_h$ of the comet and time step $\Delta t$ of the computation are calculated as follow:
\begin{eqnarray}
r_h & = & a(1-e~\cos{x_i}) \qquad (AU) \\
\Delta t & = &\frac{a^3}{G M_s} (1 - e~\cos{x_i})~ \Delta x_i \qquad (s)
\label{deltat}
\end{eqnarray}
where $a$ (AU) is the semi-major axis of the orbit of the comet, $e$ its eccentricity, $x_i$ the eccentric anomaly, $M_s$ the mass of the
star (assumed to be solar) and $\Delta x_i$ the angular step.  At the beginning of the computation (at aphelion), $x_i=-\pi$, $r_h=a(1+e)$, $\Delta t=\Delta t_0$ (which is a fraction of day, see \S\ref{timestep}), and $\Delta x_i =\frac{\Delta t_0}{\frac{a^3}{G M_s} (1 + e)}$. As soon
as the comet approaches the star, $r_h$ and $\Delta t$ decrease.

\section{Application to comets in exo-planetary systems \label{paramsection}}

The quasi 3D model of cometary nucleus described in \S\ref{sec:models} is used to determine the thermo-physical evolution of comets in extra-solar systems of different stellar luminosity.
As our knowledge of comets in exo-planetary systems is currently somewhat limited, we consider only their broad properties, choosing therefore solar cometary values for the parameters required in the cometary model. 

\subsection{Chemical composition\label{chemicalcompo}}
In the solar system, the chemical composition of the cometary ices is qualitatively and, in many cases, quantitatively consistent with that determined for the major components of astronomical ices and gas (Mumma 1997; Irvine et al. 2000; Langer et al. 2000; Gibb 2000; Mumma \& Charnley 2011). In the ISM, as in comets of the solar system, H$_2$O is the major volatile species around refractory grains (Gibb et al. 2000; Boogert \& Ehrenfreund 2004; Dartois 2009). 
Although extrasolar comets could be made up of a large proportion of volatile species such as CO and CO$_2$\footnote{The molar abundance of H$_2$O molecules could represent between 55\% and 75\% of chemical species included in icy planetesimals and further comets (see Marboeuf et al. 2014).} (Bockel{\'e}e-Morvan et al. 2004; Mumma \& Charnley 2011), we consider only H$_2$O ice as chemical species in comets of this study in order to avoid excessively long computing time. 
As H$_2$O is the most refractory element of volatile molecules, the physical erosion that comets could suffer during
their traveling should be reduced compared to comets that incorporate a lot of highly volatile species such as CO. Consequently, the interior of the comet is modeled by a homogeneous composition made of water ice and dust grains at the beginning of the computation.

\subsection{Dust to ice mass ratio and volumes}
The dust/ice mass ratio $J_{dust}$ is assumed to be equal to 1. This is the value indicated for the comet 1P/Halley by Giotto-DIDSY measurements (McDonnell et al. 1987), prescribed by Greenberg's (1982) interstellar dust model (Tancredi et al. 1994) and given by Lodders (2003) for solar system and
photospheric compositions. This value corresponds also to the minimum one for the nucleus of the comet 67P/CG (Kofman et al. 2015). However, this value could change in comets, depending probably on their area of formation in the stellar system (Marboeuf et al. (2014). The change in the dust production rates depending on this parameter can be readily estimated by retaining the sublimation rates for ices, and multiplying by the appropriate factor ($J_{dust}$) to determine the dust production rate.

The size distribution of dust grains in comets follows a power law of order $-3.5$ (McDonnell et al. 1986; Huebner et al. 2006) with a
cut-off at radii of 10$^{-6}$ m (minimum value) and 1 cm (maximum value, Prialnik 1997). We note however than the size distribution does not change the total mass of dust ejected in space since all dust grains freed by the sublimation of water ice are fully ejected in space from the surface of the nucleus (see Sec.~\ref{dustejection}). This approximation is in good agreement with the hypothesis of a dust ejection proportional to the gas flux (Horanyi et al. 1984; Prialnik et al. 2004). It remains fully valid near the star where sublimation of ice is the largest but can provide erroneous higher dust productions at largest distances where the sublimation of ice is low (low surface temperature). 
However, using this approximation, our calculations showed that the mass of dust produced far away from the star (where temperature is low) remains negligible compared to the total mass lost on the orbit (from about 2\% to 0.00005\% at 0.1 and 70 $L_\odot$ respectively).

Even though silicates should be a prominent component of dust grains of debris disks, little is known about their composition (Moro-Martin 2013).
The bulk density of dust grains, considered as a mixture of silicates (density of 3.5 g cm$^{-3}$, see Lebreton et al. 2012) and organics compounds (density of $\approx$2 g cm$^{-3}$, see Lebreton et al. 2012) in this study, is assumed to be 3 g cm$^{-3}$ (Marboeuf et al. 2012, Huebner et al. 2006). By imposing a dust/ice mass ratio $J_{dust}$ equal to 1 in comets, the volume of H$_2$O ice (bulk density of 917 kg m$^{-3}$, Feistel \& Wagner 2006) will be always about three times higher than the dust one.
This implies that the volumes of H$_2$O ice and dust grains represent respectively about 75\% and 25\% of the bulk volume of cometary material. This ratio is in good agreement with the chemical composition adopted by Lebreton et al. (2012) to reproduce the spectral observations of disks by Hershel. Indeed, the values adopted by these authors correspond to volume ratios of about 65 for ices and 35\% for refractory compounds.

\subsection{Density and thermodynamics parameters}
Estimates for the density of comets, obtained from observations and space missions in solar system, range between 100 and 1000 kg m$^{-3}$
(Davidsson et al. 2007, 2009; Sosa \& Fern\'andez 2009; Richardson et al. 2007; Lamy et al. 2007;
Davidsson $\&$ Gutierrez 2004, 2005, 2006; Davidsson \& Skorov 2002). These
values, coupled with size measurements, lead to porosities equal or greater than 50\%. In this work, we have chosen porosities of 50, 60 and 70\% leading to bulk densities of about 700, 560 and 420 kg m$^{-3}$, respectively. We note that the two last densities correspond approximately to the estimation of the 67P/Churyumov-Gerasimenko comet one (about 470$\pm$45 kg m$^{-3}$, see Sierks et al. 2015).

An important unknown parameter is the thermal inertia, i.e. the thermal conductivity, of the porous icy matrix of comets. The thermal inertia of the surface of a comet can vary of several orders of magnitude between 40 and 3000~W m$^{-2}$ K$^{-1}$ s$^{\frac{1}{2}}$ (Davidsson et al. 2009) depending on the type of surface. A low/high thermal inertia increases/decreases the temperature of the surface of the comet, leading respectively to an increase/decrease in the rate of sublimation of water ice. The rate of production of gas and dust grains from the nucleus could therefore be quite different between these two extremes values.  In this work, we test heat conductivities of about 1, 10$^{-2}$ and 10$^{-3}$~W m$^{-1}$ K$^{-1}$ using Russel formula, i.e. high thermal inertia, and Hertz factor (values of 10$^{-2}$ and 10$^{-3}$), i.e. low thermal inertia (see Marboeuf et al. 2012 for more details). For solar luminosity, this leads to a thermal inertia (by using the relation $I = \sqrt{\rho c K^m}$, where $\rho$ is the mass density (kg m$^{-3}$), $c$ the mass heat capacity (J kg$^{-1}$ K$^{-1}$), and $K^m$ the thermal conductivity of the material (W m$^{-1}$ K$^{-1}$); see Davidsson et al. 2009) of the nucleus of about 1000 W m$^{-2}$ K$^{-1}$ s$^{\frac{1}{2}}$ (high value) and of about 50 and 20~W m$^{-2}$ K$^{-1}$ s$^{\frac{1}{2}}$ (lower values). Finally, we define three models of comets named high, nominal, and low models using heat conductivities and porosities (see Table \ref{paramss}) allowing to frame extreme thermal behaviors of comets in stellar systems. The nominal model simulates a comet with a porosity of 60 \% and a thermal inertia of about 50~W m$^{-2}$ K$^{-1}$ s$^{\frac{1}{2}}$. The "low" model represents a comet with lower porosity (50 \%) and higher thermal conductivity, i.e. thermal inertia. The "high" model simulates a high porous model (70 \%) with lower thermal inertia.
 
Other thermodynamics parameters such as equilibrium pressure, enthalpies of sublimation, bulk densities of cometary materials such as ices and dust, thermal conductivities and heat capacities of solids are given and fully explained in Marboeuf et al. (2012).  
The initial temperature is assumed to be 30 K.

\subsection{Dust mantle \label{dustmantle}}

The formation of a dust mantle at the surface of the nucleus
  is mainly a function of 1) the size of the grains initially embedded
  in the water ice and 2) the flow of gas escaping from the comet at a
  given latitude and longitude, which depends on the distance to the
  star and on stellar luminosity. It depends also on 3) the size of
  the nucleus, 4) its density, i.e. its mass, and 5) its rotational period. Finally, the
  dynamical/thermodynamic history of the comet is important for the
  process of dust mantle formation.

 As described in Marboeuf \& Schmitt (2014), the presence of a
   dust mantle at the surface of the nucleus reduces greatly the
   outgassing rate of water, and consequently the dust production
   rate. This reduction depends greatly on the dust mantle thickness
   which cannot readily be estimated without a good knowledge on the
   process of dust mantle formation, and the dynamical/thermodynamic
   history of the comet.  For these reasons and to keep the problem
   tractable, we suppress the dust mantling formation process in this
   study: all dust grains freed by sublimation of water ice are then
   fully ejected from the surface for each facet of the
   comet. Therefore, the results presented in this study should be
   regarded as the maximum production rates suffered by the bodies,
   the lower production rate being simply zero, which corresponds to
   inactive comets without tails very much resembling to asteroids.

\subsection{Orbital parameters and stellar luminosity}
The three models (high, nominal, and low) are used to determine the thermo-physical, water outgassing and dust ejection behavior of a
comet on any orbit.
The orbit of a comet is important since it determines its thermodynamic, its outgassing and dust ejection behavior, and hence the erosion of the comet. The closer a comet approaches the star, the more physical changes it suffers. The pericenter (i.e. orbit) of a comet is then very important since this position corresponds approximately to the maximum erosion (maximum rate of H$_2$O sublimation) that a comet can undergo during its travel around the star.
The orbit of comets in the solar system are mainly dependent of the gravitational interactions with giant planets, in particular with Jupiter and Neptune planets (Jupiter family comets, Halley type comets, long period comets; Levison et al. 2001, Dones et al. 2004, Lowry et al. 2008). It is then impossible to reproduce these orbits in any stellar system since masses and positions of planets differ from the solar system.
Since it is impossible to provide a typical orbit of comet in any stellar system, we have arbitrarily adopted orbital parameters (see Table \ref{paramss}) of a fictive comet orbiting around one star with a single aphelion position (50 AU) and 3 possible pericenter positions: 0.1 AU, 0.5 AU, and 0.9 AU (hereafter orbits A, B, and C respectively).
The three orbits were chosen to have perihelion position within 1 AU radius around the star where exo-zodiacal dust is frequently observed (Absil et al. 2013, Ertel et al. 2014). 
The choice of these orbits allow us to draw the thermal evolution, physical changes, and the maximum mass productions of water gas and dust from comets in a wide range of stellar distances, from 0.1 AU to 50 AU. 
We assume that the comets always remain on the same orbit: orbital changes due to gravitational interactions with planets are not taken into account in this study.

For each orbital parameter and comet model, we use stellar luminosities varying from 0.1 to 70 times the solar luminosity ($L_\odot$). We assume in this work that the luminosity of the star is homogeneous on its surface.
The minimum (0.1 $L_\odot$) and maximum (70 $L_\odot$) stellar luminosities exceed slightly below and above those of the dwarf star HD 69830 (0.45 $L_\odot$) and Vega star (37$\pm$3 $L_\odot$ in average\footnote{The luminosity of this star varies between 28$^{+8}_{-6}$ $L_\odot$ and 57$\pm$3 $L_\odot$, depending whether the star is viewed pole or edge-on (Aufdenberg et al. 2006).}) respectively (see Sect.~\ref{disks}), where hot/warm exozodiacal dust disks have been observed (Beichman et al. 2005; Absil et al. 2006).

\subsection{Radius, rotational period and geometric albedo}
The comet's radius adopted in this work is 10~km. Although superior to that of JFCs hovering around 2-3 km (see Fern{\'a}ndez et al. 2013; Weiler et al. 2011; Snodgrass et al. 2011; Tancredi et al. 2006), this size corresponds to an intermediate size of comets in the solar system\footnote{Comet's radii vary from about 1 to 30$\pm$10 km, the radius of the Hale-Bopp comet (Fernandez, 2002)} and allows us to study comets around stars of very high stellar luminosity: the higher the luminosity, the higher the erosion of the comet, reducing then the lifetime of smaller objects (see \S\ref{lifetimecomets}). The obliquity of the comet is arbitrarily chosen to 0$^{\circ}$ and its rotational period is arbitrarily chosen to 12~h which is of the same order of magnitude that those of the long period comet Hale-Bopp (11h46$\pm$0h25, Warell et al. 1999), the Jupiter Family Comet (JFC) 67P/Churyumov-Gerasimenko (12h76 to 12h40, Lowry et al. 2012; Mottola et al. 2014), and the Encke-type comet 2P/Encke (11h, Lowry \& Weissman 2007) in the solar system. We note however that comets can have periods of rotation much lower such as the 107P/Wilson-Harrington comet (3h57, Urakawa et al. 2011) or much greater such as the JFC 9P/Tempel 1 (41h, A'Hearn et al. 2005). The geometric albedo adopted for the nucleus is 0.04 as for the comet C/1995 O1 Hale-Bopp in the solar system (Lamy et al. 2004). This is a intermediate value between the minimum (0.02) and maximum (0.06) values of comets (Lamy et al. 2004).

\subsection{Numerical parameters: time step and thickness of layers \label{timestep}}

At the beginning of the computation (aphelion), the time step $\Delta$$t_0$ is deliberately chosen as a small fraction of the day, i.e. 0.5\% (7 mn.).
It decreases up to the perihelion passage to about 1 s. This small time step prevents large temporal temperature variations and erosion on the surface of the nucleus during one step, especially near the star, and allows us to study the rotation effects of the comet on the temperature and phase changes of H$_2$O at the surface.
In order to better study the physico-chemical changes near the cometary surface, the first 20 (from the surface) radial layer thickness are deliberately inferior to the thermal skin depth $L$ ($L=\sqrt{\frac{K^m}{\rho~c~\omega}}$ with $\omega=\frac{2\pi}{P_r}$, where $P_r$ is the nucleus rotation period of the comet (s); see Prialnik et al. 2004, Davidsson et al. 2013) of the diurnal heat wave (see \S\ref{lifetimecomets}) for each section ($\varphi$, $\theta$) of the nucleus.
For the nominal model, the diurnal skin depth reaches about 3 cm. For low and high models, it reaches about 1 cm and 30 cm respectively.
Whatever the models, we adopt the smaller thickness (3$\times$10$^{-3}$ m) for the first 20 layers from the surface of the comet (see Tab.~\ref{paramss}).
At the end of each time step, after erosion, the size and the number of cells are recalculated in order to keep small thicknesses of layers near the surface, where all major physico-chemical changes occur (see Marboeuf et al. 2012).

\begin{table*}
\centering 
\caption{Initial parameters of the comet nucleus.}
\begin{tabular}{llcccc}
\hline
Parameter		& Name & unit										& \multicolumn {3} {c}{Value}				\\ 
\hline
Orbital parameters & \multicolumn {5} {c}{} \\
    &       Orbit& 		 &				A	&		B		&		C \\	 
$q$ &  Perihelion				&		AU			&   	0.1 &  0.5 &  0.9			\\
$Q$ &  Aphelion					&	  AU			&	    \multicolumn {3} {c}{50}\\	
$P$ & Period of revolution & years &    125  & 126.86 & 128.4 \\
$R$ & Radius  & km 									& \multicolumn {3} {c}{10} 	 		\\
$P_r$ & Rotational period & h						& \multicolumn {3} {c}{12$^{(a)}$} 				\\
& Obliquity &  &    \multicolumn {3} {c}{0} \\
\hline
Physical parameters & \multicolumn {5} {c}{} \\
 & Models &                &   high & nominal & low \\   
$\Psi^i$  & Initial porosities &    \% &		70	&    60    & 50		\\
$\rho$    & Initial density    & kg m$^{-3}$ & 420 & 560 & 700 \\ 
  &        Heat conductivity & W m$^{-1}$ K$^{-1}$  &    $\approx$10$^{-3}$&   $\approx$10$^{-2}$ & $\approx$1 \\
$T$ & Initial temperature & K 							&  \multicolumn {3} {c}{30} 				\\
$\epsilon$ & Infrared surface emissivity  & 									& \multicolumn {3} {c}{1}					\\
$A_l$ & Bolometric Bond's Albedo  & 										&  \multicolumn {3} {c}{0.04}				\\
$\tau$ & Tortuosity &   							& \multicolumn {3} {c}{$\sqrt{2}$$^{(b, c)}$} 				\\
$r_p$  & Average pore radius 	&  m						& \multicolumn {3} {c}{10$^{-5}$} 				\\
$\beta$ & power law size distribution &  & \multicolumn {3} {c}{-3.5$^{(d)}$}	\\
        & of dust grains              &  &  \multicolumn {3} {c}{}  \\
$J_{dust}$ & Dust/Ice mass ratio & 					&  \multicolumn {3} {c}{1}  \\ 
$L_\odot$  &  Solar luminosity    &   (W m$^{-2}$)  & \multicolumn {3} {c}{1360}\\
$L_{\star}$  &  Stellar luminosity    &   $L_{\odot}$  & \multicolumn {3} {c}{0.1 - 70}\\  
\hline
Numerical parameters & \multicolumn {5} {c}{} \\
$\Delta r$  & Thickness of first layers &   m  &       \multicolumn {3} {c}{3 10$^{-3}$} \\  
$\Delta t_0$  & Time step at aphelion &   day  &       \multicolumn {3} {c}{0.5\%} \\  
\hline
\end{tabular}
\label{paramss}

$^{(a)}$Warell et al. (1999), Lowry et al. (2007, 2012); $^{(b)}$Kossacki \& Szutowicz (2006); $^{(c)}$Carman (1956), Mekler et al. (1990); $^{(d)}$ McDonnell et al. (1986), Huebner et al. (2006)
\end{table*}

\section{Results \label{results}}

We investigate the thermo-physical evolution of comets around stars of
luminosities varying from 0.1 to 70~$L_\odot$. We discuss the erosion,
the mass loss in space, and the lifetime of such bodies in a broad
range of stellar environments.
The sublimation of water ice at the surface of comets, and the
resulting erosion of bodies, is independent of their size. The erosion
of comets is mainly function of the temperature of their surface which
depends on the distance to the star, latitude of sections of the
cometary surface, and on the stellar luminosity.
Therefore, we provide in this study mass production rates per unit
surface area of cometary nucleus at any position around a single star
which allow to readily derive the total mass of material ejected in
vacuum space by any comet along its orbit. The production
  rates, erosion, and lifetimes of the comets discussed in the
  following have been calculated fairly accurately by taking into
  account all facets (latitudes and longitudes) of the cometary
  nucleus, but these are averages over the cometary surface to
  facilitate their usage in other studies through empirical laws that
  we provide in the next sections.  
Moreover, we note that the cometary nucleus surface considered in
the study, to calculate physical quantities per unit of surface area,
takes into account both day and night sides of the comet even if the
night side does not suffer of water evaporation and dust
ejection. Calculations have been performed for the low, nominal, and
high models.

\subsection{Erosion of comets \label{cometerosion}}
\begin{figure}
\includegraphics[width=9.cm]{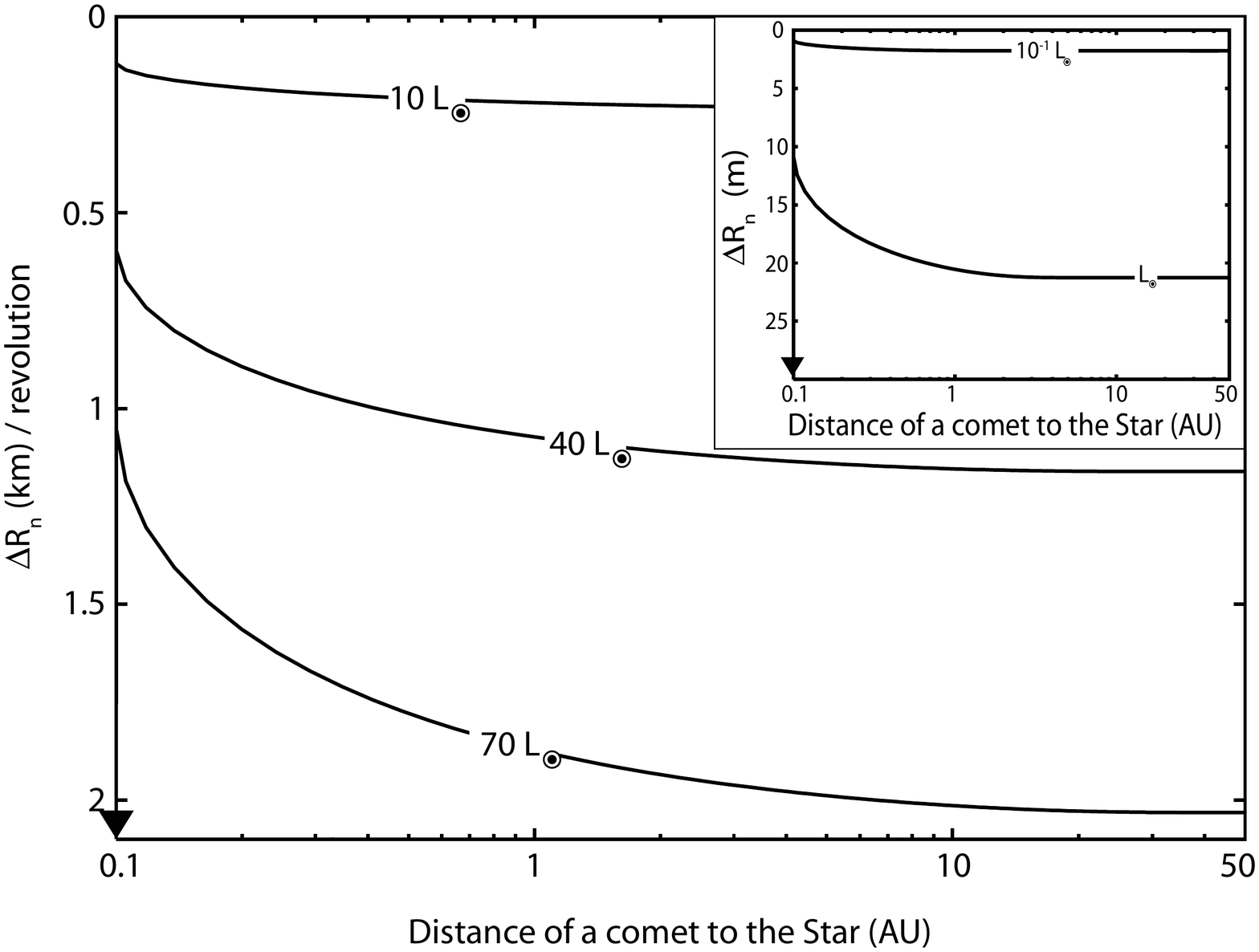}
\caption{Average thickness of matter removed from a comet, during one revolution, as a function of the distance to the star, and for several stellar luminosities. Calculations have been performed for the nominal model and orbit A.}
\label{fig:ablation_evolution}
\end{figure}

The erosion of a comet depends mainly on the temperature, i.e. the distance to the star and its luminosity.
Figure \ref{fig:ablation_evolution} presents the average thickness\footnote{The average erosion of the comet has been calculated by taking into account the erosion of all facets (longitudes and latitudes) of the cometary nucleus \label{noteaveragethickness}} of cometary material removed from the nucleus per revolution as a function of the distance to the star (from 0.1 to 50 AU, orbit A in this example), and for the nominal model. Results are presented for stellar luminosities of 10$^{-1}$ and 1 (small panel in the upper right of the figure), 10, 40, and 70 $L_{\odot}$ (large panel).  For stellar luminosities of 10$^{-1}$ and 1 $L_{\odot}$, the comet undergoes sublimation respectively up to 0.2 and 2 AU. The average thickness removed per revolution reaches some meters with respectively 2 and $\approx$ 20 m for luminosities of 10$^{-1}$ and 1 $L_{\odot}$.  For larger stellar luminosities, the erosion can reach several hundred of meters and up to
$\approx$ 2 km for simulations with 70 $L_{\odot}$. With this luminosity, the comet undergoes physical alteration up to 10-20 AU. Such thicknesses of erosion for high stellar luminosities far from the central star exceed by several orders of magnitude the physical alteration encountered by comets in the solar system.

\subsubsection{Erosion per revolution as a function of the stellar luminosity \label{erosionluminosity}}
\begin{figure}
\includegraphics[width=9.cm]{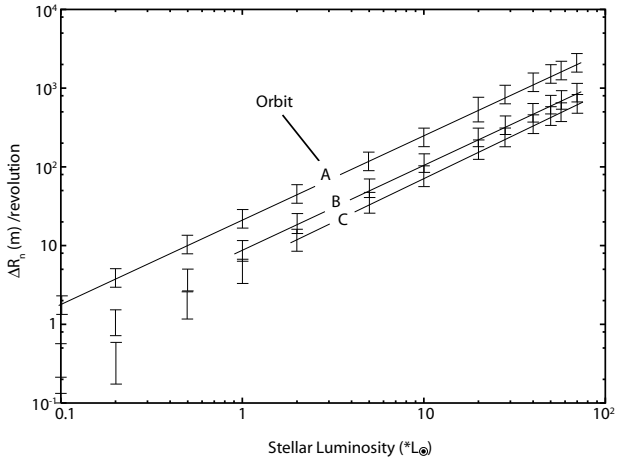}
\caption{Average thickness $\Delta R_n$ (m) of cometary material removed at
  the surface of the nucleus per revolution as a function of the stellar luminosity
  $L_{\star}$ (in $L_{\odot}$). Calculations have been performed for orbits
  A, B and C. Dots represent the data for the nominal model. Data of
  "high" and "low" models are represented respectively by high and low
  values of error bars. Line curves are fitted by
  Eq.~(\ref{eq:fit:lambda}).}
\label{fig:ablation}
\end{figure}

In this part, calculations of the physical alteration of comets have been performed for several orbits (A, B, and C), porosities and thermal conductivities of cometary materials (low, nominal, and high models).
Figure~\ref{fig:ablation} presents results of computations for the average thickness $\Delta R_n$ (m) of material removed$^{\ref{noteaveragethickness}}$ from the surface of comets per revolution as a function of the stellar luminosity, and for all models. Dots represent the values for the nominal model. Values for high and low models are respectively represented by high and low values of error bars. 

Erosion increases with decreasing pericenter distances and increasing porosities. For solar luminosity, comets suffer an erosion of $\approx$ 20, 10 and 5 meters for orbits A, B and C respectively, and for the nominal model. For higher stellar luminosity, the thickness of
cometary material removed from the cometary surface can reach up to 2 orders of magnitude the one of comets in the solar system. For 70 $L_{\odot}$, the erosion of the comet exceeds 2.5 km per revolution for orbit A and high model but drops to approximately 0.9 and 0.6 km respectively for orbits B and C, and nominal model.
From these results, we have determined an average thickness $\Delta R_n$ of cometary material removed per revolution as a function of the stellar luminosity by the following power law:
\begin{equation}
\Delta R_n (L_{\star})= \nu  L_{\star}^{\mu}  \qquad (m) ~/~ revolution
\label{eq:fit:lambda}
\end{equation}
where $L_{\star}$ is expressed in $L_{\odot}$, and $\nu$ and $\mu$ are
given in Table~\ref{param5} for each orbit. The dependence on the
luminosity is almost linear ($\mu$ close is to 1).
\begin{table}
\centering 
\caption{Parameters allowing the calculation of the erosion $\Delta
  R$ (m) (Eq.~\ref{eq:fit:lambda}) and lifetime (Eq.~\ref{eq:fit:lifetime2}) of comets, both for orbits A, B, and C, and all models.}
\begin{tabular}{lccc}
\hline
Orbit & $\nu$ (m) 			 & $\mu$   					  & luminosity range ($L_{\odot}$) \\ 
\hline
   A	& 21.47 $\pm$ 1.5     & 1.08 $\pm$ 0.026   &    0.1-70 \\
   B  &   8.96 $\pm$ 1.26   &  1.086 $\pm$ 0.046 & 1-70 \\
   C  &  5.61 $\pm$ 1.11    & 1.12 $\pm$ 0.062   &  2-70 \\
  \hline
\end{tabular}
\label{param5}
\end{table}
The thickness of cometary material removed depends on its density and thermal conductivity, orbital parameters of the comet
and luminosity of the star. The figure~\ref{fig:ablation} and Eq.~(\ref{eq:fit:lambda}) are therefore only valid for orbits A, B and C. By changing orbital parameters, the thickness of cometary material removed from the nucleus can change of several tens of \% per revolution. However, Fig.~\ref{fig:ablation} and Eq.~(\ref{eq:fit:lambda}) give a good order a magnitude of the physical alteration experienced by comets per revolution in extra-solar systems, and assuming that dust grains are fully ejected in space. Moreover, the thickness of material removed at the surface of comets is directly proportional to the flow of water (see Eq.~\ref{erosion_comet_water}), which is itself almost proportional to the Luminosity (see Eq.~\ref{boundary_heat_surf}). It results that the thickness (Eq.~\ref{eq:fit:lambda}) and mass (Eq.~\ref{eq:fit:sigma}) removed at the surface of the comet are almost proportional to the stellar luminosity.

\subsection{Dust and water gas released by comets}
\begin{figure}
\includegraphics[width=9.cm]{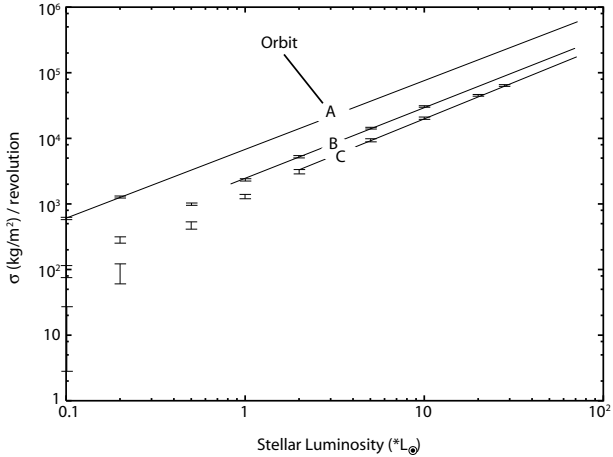}
\caption{Total mass of H$_2$O molecules (resp. dust) ejected per unit
  of surface area of the comet (kg/m$^2$) per revolution as a function
  of the stellar luminosity. Calculations have been performed for
  orbits A, B and C, and for all models. Dots represent the values of the nominal
  model. Values of high and low models are respectively
  represented (if they are different from the nominal model) by high
  and low values of error bars. Line curves are fitted by
  Eq.~(\ref{eq:fit:sigma}).}
\label{fig:masslost}
\end{figure}

The surface erosion of comets described in Sect.~\ref{erosionluminosity} is a consequence of the sublimation of water ice, and the resulting water outgassing and dust ejection from the surface.
Figure~\ref{fig:masslost} presents the {\sl total} mass $\sigma$
(kg m$^{-2}$) of H$_2$O molecules (respectively dust, $J_{dust}=1$)
ejected in space per unit of cometary surface area\footnote{We
    note that $\sigma$ has been calculated by taking into account
    flows of gas and dust escaping from all facets (latitudes and
    longitudes) of the nucleus, and divided by the comet surface in order to produce average data.} and per revolution as a function of the stellar luminosity $L_{\star}$, for the three models. 
$\sigma$ varies by several orders of magnitude from 10 to 10$^6$~kg m$^{-2}$ for stellar luminosities varying from 0.1 to 70~$L_\odot$ whatever the orbital
parameters used in the study.
From these results, we have determined that $\sigma$ follows a power law with the stellar luminosity $L_{\star}$ for each orbit:
\begin{equation}
\sigma(L_{\star})=\alpha  L_{\star}^{\beta}  \qquad (kg \ m^{-2}) ~/~ revolution
\label{eq:fit:sigma}
\end{equation}
where $L_{\star}$ is expressed in $L_{\odot}$, and $\alpha$ and $\beta$ are
given in Table~\ref{param6} for the three orbits. 
\begin{table}
\centering 
\caption{Parameters allowing the calculation of the total mass
  released (see Eq.~\ref{eq:fit:sigma}) per unit of surface area of a
  comet and per revolution $\sigma$ (kg m$^{-2}$), both for orbits A,
  B, and C, and all models.}
\begin{tabular}{lccc}
\hline
Orbit & $\alpha$ (kg m$^{-2}$) & $\beta$  						& Luminosity range ($L_{\odot}$)\\ 
\hline
  A	&  7050.83 $\pm$ 68.5			 & 1.044 $\pm$ 0.004 		& 0.1-70 \\
  B &  2512.93 $\pm$  40.4		 & 1.082 $\pm$ 0.005		& 1-70 \\
  C &  1601.11 $\pm$ 37.94		 & 1.11 $\pm$ 0.007			& 2-70 \\
 \hline
\end{tabular}
\label{param6}
\end{table}
The dependence on the luminosity is almost linear ($\beta$ is close to 1). Only values for stellar luminosities lower than or equal to 1 $L_{\odot}$ show discrepancies with the power law equation (\ref{eq:fit:sigma}) and only for the farest perihelion positions (orbits B and C). In all other cases, the total mass released varies almost linearly with the stellar luminosity. 

For a given stellar luminosity and orbit, the mass of H$_2$O molecules (resp. dust) ejected by comets remains approximately constant whatever the porosity and thermal conductivity of the cometary material. Only the thickness of cometary material changes with models as shown by Fig.~\ref{fig:ablation}. The
mass lost per unit of surface area of comets and per revolution is then mainly function of the stellar luminosity and orbital
parameters. We note that the figure~\ref{fig:masslost} is therefore only valid for orbits A, B and C. By changing these orbits, the mass of H$_2$O molecules (resp. dust) ejected in space would change by several tens of \% per revolution.

\begin{figure}
\includegraphics[width=9.cm]{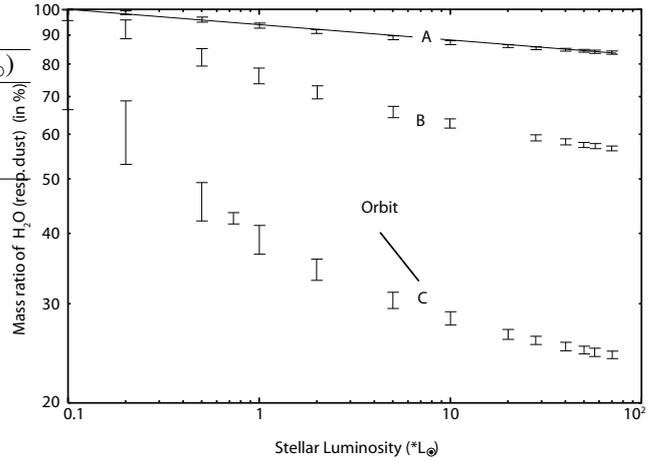}
\caption{Mass ratio of H$_2$O molecules (resp. dust) ejected by the
  comet within 1 AU from the star relative to the total mass
  of H$_2$O molecules (resp. dust) ejected per revolution as a
  function of the stellar luminosity. Calculations were carried out
  for orbits A, B and C, and for all models. Dots represent the values of the nominal
  model. Values of "high" and "low" models are represented
  respectively by high and low values of error bars.}
\label{fig:ratio}
\end{figure}

Since exozodiacal dust is frequently observed in an area of 1 AU around stars (see Absil et al. 2013; Ertel et al. 2014), we investigate the mass fraction of gas H$_2$O molecules
(resp. dust) ejected by the comet in this area relative to the total mass of H$_2$O molecules (resp. dust) ejected
per revolution as a function of the stellar luminosity, and for all models (see Fig.~\ref{fig:ratio}). 
Even if dust grains fall onto the star or are ejected outside of this area due to radiation pressure, it is interesting to determine the ratio of mass of dust and water gas ejected by a comet orbiting near the star for further comparisons with observations.
First, the mass of cometary material released far away from the star increases with stellar luminosity. It results that the relative (to total mass ejected) mass fraction of water (resp. dust) released in the area of 1 AU decreases with the increasing stellar luminosity. For orbit A, the behavior follows a quasi power law with the
stellar luminosity. 
More than 80\% of the mass ejected by comets per revolution is lost within 1 AU from the star whatever the stellar luminosity. By increasing the perihelion position (orbits B and C), the mass fraction of H$_2$O molecules (resp. dust) ejected by the comet decreases faster with increasing stellar luminosity. If the minimum ratio represents respectively $\approx$ 95\% and 70\% for orbits B and C, and for the stellar luminosity 10$^{-1}$ $L_{\odot}$, it
decreases respectively to $\approx$ 56\% and 25\% for 70 $L_{\odot}$.
From these results, the contribution of a comet in the area of 1 AU is maximum for perihelion positions lower than or equal to 0.5 AU. The higher is the stellar luminosity, the higher the erosion of comets outside of the area of 1 AU around the star. For such high stellar luminosity, comets can actively contribute to populating the inner regions of extrasolar systems with dust.

 \subsection{Lifetime of comets \label{lifetimecomets}}

A comet cannot survive indefinitely. It has a finite lifetime based on
the rate of cometary material removed per revolution and its initial
size.  Physical alteration of comets presented in the previous
subsection should decrease considerably the lifetime of these
bodies. If we assume that comets do not change orbit, i.e. pericenter
position, during their lifetime, one can derive an order of magnitude of the number of revolutions $N_r$ around stars and the lifetime $\tau$ of comets per unit of comet's radius before they fully disappear by erosion, by using the average erosion $\Delta R_n(L_{\star})$ of comets (see Eq.~\ref{eq:fit:lambda}):
\begin{equation}
N_r(L_{\star})= \frac{10^3}{\Delta R_n(L_{\star})} = \frac{10^3}{\nu L_{\star}^\mu} \qquad (km^{-1})
\label{eq:fit:lifetime1}
\end{equation}
and
\begin{equation}
\tau(L_{\star})= P~N_r(L_{\star}) = \frac{10^3 \times P}{\nu L_{\star}^\mu}  \qquad (years~km^{-1})
\label{eq:fit:lifetime2}
\end{equation}
where $\Delta R_n(L_{\star})$ is the average erosion (in meters) of the comet, and $P$ the period of revolution of the comet around the star (in years). $L_{\star}$ is expressed in $L_{\odot}$, and $\nu$ and $\mu$ are given in Table~\ref{param5} for each orbit. The lifetime of a comet is mainly function of its initial size, orbit, and the luminosity of the star, i.e. erosion $\Delta R_n(L_{\star})$ of the comet.

\begin{figure}
\includegraphics[width=6.cm, angle=-90]{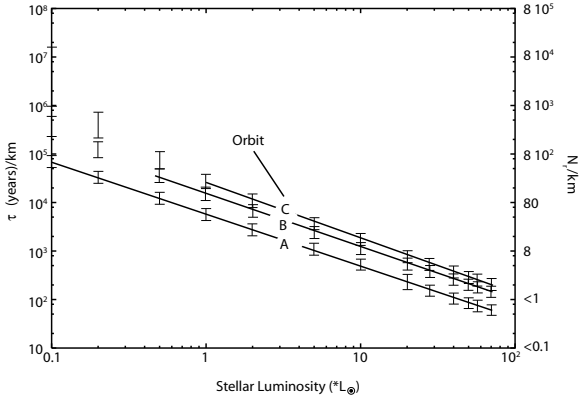}
\caption{Lifetime $\tau$ and number of revolutions $N_r$ of a comet per unit of comet's radius (in km) as a function of the stellar luminosity. Calculations were carried out for orbits A, B, and C, and all models. Dots represent the values of the nominal model. Values of "high" and "low" models are represented respectively by low and high values of error bars.}
\label{fig:lifetime}
\end{figure}

Figure~\ref{fig:lifetime} shows the lifetime $\tau$ (resp. number of revolutions $N_r$ around stars) of a comet per unit of comet's radius (in km) as a function of the stellar luminosity.
Calculations have been performed for orbits A, B, and C, and low, nominal, and high models.
Fig.~\ref{fig:lifetime} and Eq.~(\ref{eq:fit:lifetime1}) show that the lifetime of comets (resp. $N_r$) decreases with increasing luminosity and closer perihelion position to the star. The dependence on the luminosity (resp. $N_r$) is almost inversely proportional to $L_{\star}$ ($\mu$ close to 1, see Tab.~\ref{param5} and \S\ref{erosionluminosity}) and varies linearly with the radius of the comet. As for erosion (see Fig.~\ref{fig:ablation}), only values for stellar luminosities lower or equal to 1~$L_{\odot}$ show discrepancies with the power law equation (\ref{eq:fit:lifetime2}) and only for orbits B and C. 

The lifetime (resp. $N_r$) of a 10 km radius type comet reaches approximately 10$^6$~years (resp. $8\times 10^3$
laps) around stars of 0.1~$L_{\odot}$, using orbit A, and exceeds 10$^8$~years (resp. $8\times 10^5$ laps)
by using orbit C (see Fig.\ref{fig:lifetime}). For comets in the solar system, the lifetime of bodies varies between
$\approx 6\times 10^4$~years (resp. 480 laps) on orbit A and $\approx 6\times 10^5$~years (resp. 4800 laps) on orbit C. For the higher luminosity 70~$L_{\odot}$, their lifetime is reduced to $\approx 5\times 10^2$ years (resp. 4 laps) on orbit A and to 5 10$^3$~years (resp. 40 laps) on orbit C.

We note that by changing the initial size of the comet, we change the lifetime $\tau$ of the nucleus by the same order of magnitude because the lifetime is linearly proportional to the radius. So, for stellar systems with the higher luminosity of 70~$L_{\odot}$, $\tau$ varies from
several tens of years ($\leq$ 1 lap) for bodies of 1~km radius to more than $5\times10^4$~years ($\geq$ 400 laps) for bodies of 100~km radius. In such high luminosity environments, small icy bodies of 1~km of radius can not survive more than 500~years whatever the orbit and model used. For larger radii
such as 10~km and 100~km, $\tau$ does not exceed $5\times 10^3$ years and $5\times 10^4$ years respectively.
From these results, it is difficult to imagine that small objects of 1 km radius, or less, with small lifetimes (resp. $N_r$) in high stellar luminosity environments could be at the origin of exozodis over a long time, except if a large reservoir of comets such as the Oort cloud or the Kuiper belt in the solar system, and gravitationally disturbed, send many bodies close to the stars over a long time scale (see Discussion in Sec.~\ref{disks}).

\subsection{Mass production rates \label{analytic}}

We provide mass productions rates, $Q$ (kg m$^{-2}$ s$^{-1}$), of H$_2$O molecules and dust grains\footnote{We remember that the mass production rates are average values at the surface of the comet, considering all facets (longitudes and latitudes) of the cometary nucleus.} per unit of cometary surface area as a function of distances to the star, stellar luminosity and temperature, and for the nominal model. This is meant to be used as simplified recipes for further modeling of the contribution of comet evaporation to dust replenishment in extra-solar planetary systems.

\begin{figure}
\includegraphics[width=9.cm]{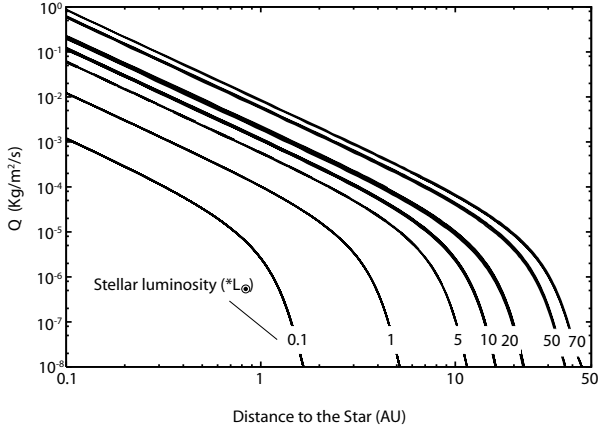}
\caption{Mass production rate $Q(r)$ (kg m$^{-2}$ s$^{-1}$) of H$_2$O molecules (resp. dust) per unit of cometary surface area and per revolution as a
  function of the distance to the star (AU), for stellar luminosities varying from 0.1 to 70~$L_{\odot}$.}
\label{fig:single:dustprod}
\end{figure}

Figure~\ref{fig:single:dustprod} diplays $Q$ as a function of the distance to the star for several stellar luminosities.
The higher the luminosity, the higher the mass production rate of
H$_2$O molecules (resp. dust) is, whatever the distance to the star.
The mass production of H$_2$O molecules (resp. dust) follows partially
a power law near the star and then drops rapidly beyond some turn-over
distance that depends on the luminosity. For the highest stellar
luminosity 70~$L_{\odot}$, the mass production rate
reaches approximately 1~kg m$^{-2}$ s$^{-1}$ at 0.1 AU perihelion. At 0.9 AU (Hale-Bopp comet perihelion position), it reaches about 5~10$^{-2}$~kg m$^{-2}$ s$^{-1}$, about 100 times more than the maximum dust mass production rate of Hale-Bopp comet in the solar system.

In order to enable easy calculation of the mass of dust and gas H$_2$O
produced by a comet of any size, on any orbit and around any star with
stellar luminosity $L_{\star}$ between 0.1 and 70 $L_{\odot}$, we
empirically fit the modeled mass loss rates by a radial power law as a
function of the distance $r$ to the star until a turn-over distance
$r_0$ beyond which the rates decrease exponentionally:

\begin{equation}
Q(r_h,~L_{\star}) = \frac{Q_0(L_{\star})}{r_h^{a}} \exp\left( -\left(
\frac{r_0(L_{\star})}{r_h} \right)^{b}\right) \qquad (kg \, m^{-2} \,
s^{-1})
\label{eq:fit:dustprod}
\end{equation}
where $r_h$ is in AU, $L_{\star}$ is in $L_\odot$, $a=2.08 \pm 0.03$,
$b=-3.03 \pm 0.08$. $Q_0(L_{\star})$ and $r_0(L_{\star})$ are only
function of the stellar luminosity $L_{\star}$ following the relations:
\begin{equation}
Q_0(L_{\star}) =  10^{-4.000}  L_{\star}^{1.036}   \qquad (kg \, m^{-2} \, s^{-1})
\label{eq:fit:dustprodQ}  
\end{equation}
with uncertainties on the power law indexes of the order of 0.005, and
\begin{equation}
r_0(L_{\star}) = 10^{0.468}  L_{\star}^{0.493}  \qquad (AU)
\label{eq:fit:dustprodr}
\end{equation}
with uncertainties of the order of 0.001 on the power law indexes, and
where $L_{\star}$ is in $L_\odot$.
We note that the mass production rate of water and dust follows an exponential law as the sublimation equilibrium pressure of water which is almost proportional to $exp(-A/T+B \times log(T)) = T^B \times exp(-A/T)$, where A and B are constant parameters.
Close to the star where the temperature is maximum, the exponential term approaches 1 and the production rate (Eq.~\ref{eq:fit:dustprod}) is simply proportional to $Q_0(L_{\star})/r_h^2$ \footnote{Close to the star, the major term on the right side of Eq.~(\ref{boundary_heat_surf}) is the sublimation energy term. It results that the free sublimation rate is proportional to $L_{\star}/r_h^2$}. Far from the star, all the terms of Eq.~(\ref{boundary_heat_surf}) have to be taken into account for the calculation of the temperature of the surface. It results that the temperature of the surface is function of $L_{\star}/r_h^2$ and provides therefore Eq.~(\ref{eq:fit:dustprod}) by substitution of $T$ in Eq.~(\ref{phi}).
Eq.~(\ref{eq:fit:dustprod}) remains valid whatever the
distance to the star for stellar luminosities higher that 1
$L_{\odot}$. For stellar luminosities of 0.5 and 1 $L_{\odot}$,
Eq.~(\ref{eq:fit:dustprod}) remains valid only in the range 0.1-10
AU. Otherwise, it takes the value 0. For the stellar luminosity 0.1
$L_{\odot}$, it remains valid only in the range 0.1-5
AU. Eq.~(\ref{eq:fit:dustprod}) can be integrated over any orbit, or
piece of orbit, to estimate the maximum mass of gas and dust produced by a comet
around a broad range of stellar environments.

\begin{figure}
\includegraphics[width=7.cm,angle=-90]{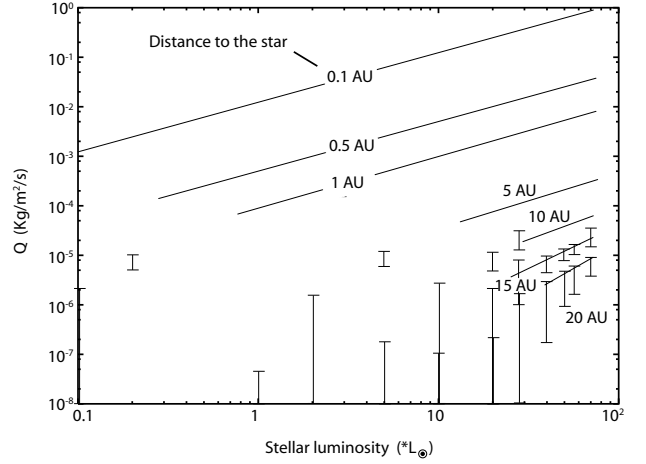}
\caption{Average mass production $Q(L_{\star})$ (kg m$^{-2}$s$^{-1}$) of H$_2$O
  molecules (resp. dust) per unit of cometary surface area and per
  second as a function of stellar luminosity (from 0.01 to
  70$L_{\odot}$). Calculations have performed for Stellar distances of
  0.1, 0.5, 1, 5, 10, 15 and 20 AU. Dots represent the values of the
  nominal model. Values of high and low models are represented
  (if they are different from the nominal model) respectively by
  high and low values of error bars.}
\label{fig:rate_Lstar}
\end{figure}

The maximum mass production rates of H$_2$O molecules (resp. dust), $Q$, can also be represented as a function of the stellar luminosity
$L_{\star}$ for different distances $r$ from the star. This is shown in Figure~\ref{fig:rate_Lstar} for $r$ = 0.1, 1, 5, 10, 15 and
20~AU. Dots represent the data for the nominal model. Data of high and low models are represented (only if different from the nominal model) respectively by high and low values of error bars. Eq.~(\ref{eq:fit:dustprod}) can be reverted, numerically for instance, to get $Q$ as a function of the star luminosity for a given
distance $r$ and for the nominal case. To complement this approach, we have fitted for each stellar distance, the mass production $Q(L_{\star})$
(kg m$^{-2}$ s$^{-1}$) of H$_2$O molecules (resp. dust) by a power law as a function of the stellar luminosity $L_{\star}$ ($L_\odot$):
\begin{equation}
Q(L_{\star})= \nu L_{\star}^{\mu}  \qquad (kg \, m^{-2} \, s^{-1})
\label{eq:fit:dustprodstar}
\end{equation}
where $\mu$ and $\nu$, given in Table \ref{param3}, are function of
the distance to the star $r$ (AU).  For $r$ greater than 1 AU, the
mass production of H$_2$O molecules (resp. dust) follows the power law
given by Eq.~(\ref{eq:fit:dustprodstar}) only for large stellar
luminosities ($L_{\star} \geq$ 20$L_\odot$).

\begin{table}
\centering 
\caption{Parameters allowing the calculation of the dust and gas mass
  production rates $Q(L_{\star})$ per unit of cometary surface area
  (kg m$^{-2}$ s$^{-1}$) as a function of the stellar luminosity
  $L_{\star}$ for several distances to the star $r$ using
  Eq.~(\ref{eq:fit:dustprodstar}). The last column indicates the valid
  range of luminosities.}
\begin{tabular}{lccc}
\hline
$r$ (AU) &  $\mu$   &  $\nu$ (kg m$^{-2}$ s$^{-1}$)  & $L_{\star}$ ($L_\odot$) \\
\hline
0.1											&		1.004 $\pm$	0.0008			& 	1.222 10$^{-2}$ $\pm$ 2.8 10$^{-5}$      &    0.1 - 70 \\
0.5											&		1.008 $\pm$	0.003			  & 	5.035 10$^{-4}$ $\pm$  4.4 10$^{-6}$      &    0.5 - 70 \\
1											 &	 	1.028 $\pm$	0.0067	  	& 	9.546 10$^{-5}$ $\pm$  1.95 10$^{-6}$      &    1 - 70 \\
5											 &    1.098 $\pm$	0.016		    & 	2.878 10$^{-6}$ $\pm$  1.7 10$^{-7}$      &    20 - 70 \\
10										 &		1.416 $\pm$	0.047		    & 	1.571 10$^{-7}$ $\pm$  2.8 10$^{-8}$      &    20 - 70 \\
15									   &		1.782 $\pm$	0.093		    & 	1.208 10$^{-8}$ $\pm$  4.45 10$^{-9}$      &    28 - 70 \\
20										 &	  2.164 $\pm$	0.12		    & 	9.13 10$^{-10}$ $\pm$  4.4 10$^{-10}$      &    40 - 70 \\
\hline
\end{tabular}
\label{param3}
\end{table}

\subsection{Comparison with solar system comets}
\begin{figure}
\includegraphics[width=9.cm,angle=0]{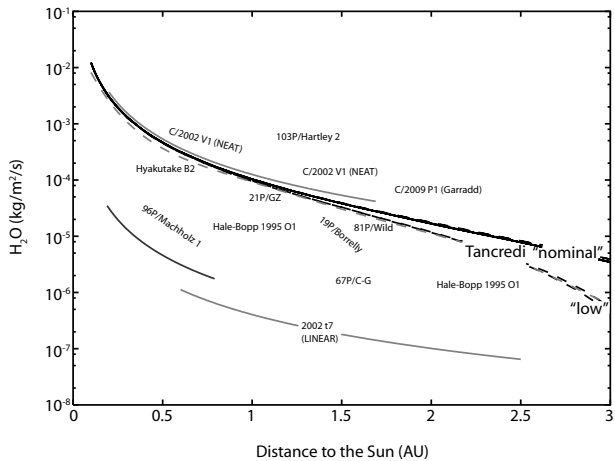}
\caption{Mass production (kg m$^{-2}$s$^{-1}$) of H$_2$O  molecules
  per unit of cometary surface area and per second as a function of
  the distance to the Sun. The dots represent the observational data
  of some comets for which sizes and H$_2$O production exist. Adapted
  from Marboeuf \& Schmitt (2014).} 
\label{fig:rateH2O_sun}
\end{figure}

\begin{figure}
\includegraphics[width=9.cm,angle=0]{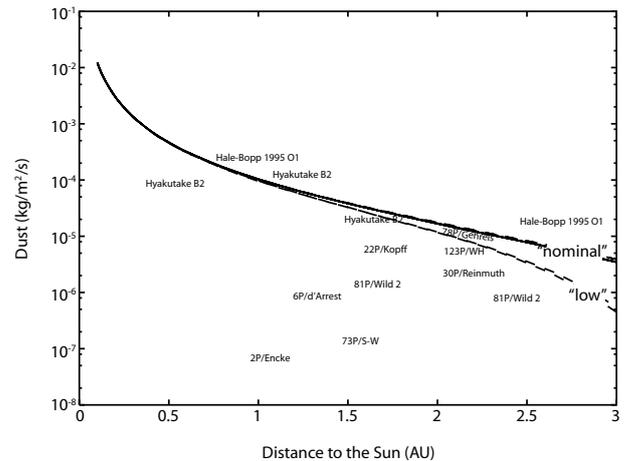}
\caption{Mass production (kg m$^{-2}$s$^{-1}$) of dust grains per unit
  of cometary surface area and per second as a function of the
  distance to the sun. Dots represent the observational data of some
  comets for which sizes and dust production exist.}
\label{fig:ratedust_sun}
\end{figure}

We here briefly compare our results with the many production
  rates of dust and H$_2$O that are available in the literature.
  Figures \ref{fig:rateH2O_sun} and \ref{fig:ratedust_sun} present
  respectively the mass production rates (kg m$^{-2}$ s$^{-1}$) of
  H$_2$O and dust per unit of cometary surface area as a function of
  the distance to the Sun (AU) for the three models (high, nominal,
  and low) and several short period and long period comets of the
  Solar System for which sizes (Lamy et al. 2004, Tancredi et
  al. 2006, Lisse et al. 2009, Sosa \& Fern\'andez 2011, Weiler et
  al. 2011, Fern\'andez et al. 2002, 2013), and H$_2$O (19P/Borrelly,
  21P/Giacobini–Zinner, 67P/Churyumov–Gerasimenko, 81P/Wild 2,
  96P/Machholz 1, 103P/Hartley 2, Hale-Bopp 1995 O1, C/1996 B2
  Hyakutake, C/2002 V1 (NEAT), 2002 t7 (LINEAR), C/2009 P1 (Garradd);
  Colom et al. 1997, Dello Russo et al. 2002, Schleicher 2006, Combi
  et al. 2009, 2011a, 2011b, 2011c, 2013) or/and dust (Hale-Bopp 1995
  O1, C/1996 B2 Hyakutake, 2P/Encke, 6P/d'Arrest, 22P/Kopff,
  30P/Reinmuth 1, 78P/Gehrels 2, 81P/Wild 2, 123P/West-Hartley
  (123P/W-H); Fulle et al. 1997, Jewitt \& Matthews 1999, Sanzovo et
  al. 2001, Lisse et al. 2002, Kidger 2003, Moreno et al. 2012)
  production rates have been calculated from observations. The black
  line corresponds to the H$_2$O production rate per unit cometary
  surface area for the nominal model. The dashed black lines represent
  the rates for the high and low models, that only diverge far from
  the Sun. The dashed gray line (only Fig.~\ref{fig:rateH2O_sun})
  represents the gas production rate of H$_2$O per unit surface area
  of fully exposed ice on the comet (averaged over the surface of a
  spherical nucleus) fitted to the average curve of Tancredi et al.'s
  simulations (Tancredi et al. 2006; see equation in Sosa \&
  Fern\'andez 2011). Other lines are the power-law fits of the
  production rates of H$_2$O to the corresponding data of the comets
  (see Combi et al. 2009, 2011a, 2011b, 2011c, 2013).

As explained in Marboeuf \& Schmitt (2014), the H$_2$O production
rates show a dispersion of several orders of magnitude from one comet
to another and compared to our theoretical data. In particular, long
period comets, which are considered as new comets (less processed and
differentiated in subsurface layers) when visiting the inner region of the solar system,
show high activities although regular comets (JFCs) show much lower
activities.
These discrepancies can be explained by the physico-chemical
heterogeneity among comets (ice-to-dust mass ratio, porosity,
thickness of the dust mantle at the surface of the comet, active
surface area fraction, obliquity, rotational period, shape), due to
different formation regions in the protoplanetary disk and different
dynamical/thermodynamical evolution, which are not considered in
this study.
Moreover, uncertainties about the physical characteristics of the
nuclei such as shape and size could also explain some of the
discrepancies (see Sect. 5 in Marboeuf \& Schmitt 2014). The high
values of H$_2$O production rate for the long-period comets can be
explained by the sublimation of icy grains ejected in the coma,
generating an additional activity which outclasses the case of a
nucleus with a 100\% active surface area (Marboeuf \& Schmitt 2014,
Combi et al. 2011b, Sosa \& Fern\'andez 2011).

Similarly to the production rate of H$_2$O, the production
  rate of dust grains can vary by several orders of magnitude from one
  comet to another (see Fig.~\ref{fig:ratedust_sun}). Our model is in
  good agreement with the Hale-Bopp and Hyakutake B2 comets which are
  considered as new comets in the inner part of the solar system, but
  present values higher than periodic comets which are probably more processed and differentiated in subsurface layers. As discussed in Combi et al. (2011b), the minimum active surface area calculated from the water production rate is typically
  between 5\% and 20\% of the physical surface area of the nuclei of
  periodic comets.
Moreover, the ratio between the dust and gas mass production rates
varies by several order of magnitude from 0.08 to 10 among comets
(Lara et al. 2004; Lisse et al. 2002; Jewitt \& Matthews 1999). This
can be explained by the comet heterogeneity (chemical composition,
ice-to-dust ratio, active surface area, ...),
dynamical/thermodynamical history of the comets, and shape/rotational
period.
We note however that the dust production rate measured for comets is
strongly model dependent, relying on assumptions about the density,
the size and velocity distribution of the dust grains (Kidger 2003;
Moreno et al. 2012).

We conclude that our results can be compared to outgassing
  rates for less processed and differentiated comets such as long period comets since our model considers a poor
  differentiated cometary nucleus, without dust mantle formation. As already explained in Sec.~\ref{dustmantle}, the absence of dust mantle formation at the surface of bodies produces a maximum outgassing and dust production rate and therefore produces the minimum number of comets needed to reproduce the observational data.

\section{Implications for (exo)zodiacal dust disks\label{disks}}

Near- or mid-infrared excess observed around many stars such as Vega
(28$^{+8}_{-6}$ L$_\odot$ to 57$\pm3$ L$_\odot$, Aufdenberg et
al. 2006, Absil et al. 2006), $\beta$ Pictoris (8.7 L$_\odot$, Crifo
et al. 1997, Defr\`ere et al. 2012), and HD 69830 (0.45 L$_\odot$, Cox
2000, Beichman 2005) is associated with warm or hot dust close to the
star. We hypothesise that a potential source or contributor to the
observed emission could be dust released as comets that approach close
to the star sublimate. Here, we briefly consider the case of the
  zodiacal dust disk, and then two examples of extrasolar systems for
  which detailed modeling of their exozodiacal dust disks exists. The
  results presented in this work can nevertheless be used to assess
  the potential contribution of exocomets to dusty emission in any
  stellar systems.

\subsection{Methodology}
In order to consider the contribution of dust released from comets, we need to consider the fate of the said dust particles. Once released by comets, the dust grains are affected by radiation and/or stellar wind pressure and drag that, depending on their size, could place the grains on very elliptical to unbound orbits, or cause them to drift inwards (Burns et al. 1979). If sufficiently dense, collisions between dust grains may be important in their evolution. In order to obtain a rough estimate of whether or not comets could contribute to observed exozodiacal dust, we estimate which of the aforementioned processes is likely to dominate the dust lifetime and use this to estimate how long the dust grains released by the comet are likely to survive in the inner regions of a planetary system. This will provide an estimate of the minimum number of comets required to produce the observed dust.

We can estimate the minimum number of comets needed to reproduce the exozodiacal dust mass around stars by assuming typical radii for exocomets and a rough estimation of the dust lifetime. The removal timescale of the grains depends on their size, the distance to the star, its mass and luminosity, and on the disk density if collisions are important for the dynamics. For example, in the solar system, zodiacal dust grains larger than $\sim 10\,\mu$m, are removed in 2-6$\times$10$^5$ years from a 1\,AU distance due to Poynting-Robertson drag,
while the much smaller, submicron-sized grains in the HD~69830 system drift inward from a 1\,AU distance in less than 1000 years, a timescale similar to the local collision timescale in this much denser system than the zodiacal cloud (Beichman et al. 2005). In the case of Vega, the situation is even worse. The star is much more luminous than the Sun or HD~69830, the submicron-sized grains are much closer to the star (0.1--0.5\,AU), and the exozodiacal disk is sufficiently dense for collisions to be important. Overall, this reduces the lifetime of the smallest grains to a year or less in the Vega exozodiacal system (Absil et al. 2006, Defr\`ere et al. 2011). Let us consider in the following a typical
1--10 km radius comet around the Sun and two extreme luminosity stars that have just been mentionned: the 450\,Myr old A-type star Vega system whose luminosity varies between 28$^{+8}_{-6}$ L$_\odot$ (equator) and 57$\pm3$ L$_\odot$ (poles, see Aufdenberg et al. 2006), and the 3--6\,Gyr old \citep{mar14}, K0V low luminosity star HD~69830 (0.45 L$_\odot$). Tables \ref{param_systems} and \ref{param_mass_dust_systems} summarise the observational data for the three systems and method used to determine the number of comets needed to reproduce the observational data.

\begin{table*}
\centering 
\caption{Summary of parameters used for the different systems.}
\begin{tabular}{|c|cccc|}
\hline
Stellar System & Luminosity (L$_\odot$) & Total mass of grains (M$_{\oplus}$) &  Lifetime of grains  (years) & Mass production rate (kg/year)$^{(*)}$\\
\hline
HD 69830      &    0.45   &   4.6$\times$10$^{-9}$$^{(a)}$     &   400--700$^{(a)}$  &  4-7$\times$10$^{13}$    \\ 
Solar         &    1      &   -        &     2-6$\times$10$^5$$^{(a)}$   &  2.7$\times$10$^{11}$$^{(b)}$   \\
Vega          &   28-57$^{(c)}$   & 10$^{-9}$$^{(d)}$   &    1$^{(e)}$   &  6$\times$10$^{15}$    \\
 \hline 
\end{tabular}
\begin{flushleft}
$^{(*)}$ The rate of production (kg/year) for HD 69830 and Vega is calculated by dividing the total mass of grains by their lifetime in the system. The one for the Solar system is given in Moro-Martin (2013).\\
$^{(a)}$ Beichman et al. (2005); $^{(b)}$ Moro-Martin (2013), the amount consider only 85\% of the production origined from comets; $^{(c)}$ Aufdenberg et al. (2006); $^{(d)}$ Defr\`ere et al. (2011); $^{(e)}$ Absil et al. (2006)\\
\end{flushleft}
\label{param_systems}
\end{table*}

\begin{table*}
\centering 
\caption{Dust mass produced and number of comets in three systems.}
\begin{tabular}{|c|ccc|cc|c|}
\hline
Column number 	&	 0 &	    I    &   II    &    III       &   IV    &   V      \\
Stellar System  & R$_n$ (km) & Area (Orbit) & Timescale$^{(a)}$ & Mass frac.$^{(b)}$ & Mass rate (kg/month)$^{(c)}$   & N$_n$ comets$^{(d)}$  \\   
  \hline   
  \hline
\multirow{2}{*}{HD 69830}  & 10 & 0.1-1 AU (A) &  2 months   &  100 \%  & 1.9$\times$10$^{12}$$^{(e)}$    &  2--3$^{(e)}$ \\ 
                           & 10 & 0.9-1 AU (C) &  1 month   & 70  \% & 8.3$\times$10$^{11}$$^{(e)}$ &   4--7$^{(e)}$  \\    
 \hline
 Solar &  2.5$^{(f)}$&     2-3 AU &  --   & -- &  1.8$\times$10$^{9}$$^{(g)}$       &  13 \\
 \hline
 \multirow{2}{*}{Vega}  &10     &  0.1-1 AU (A) & 2 months   & 85 \%&  1.2$\times$10$^{14}$--2.6$\times$10$^{14}$$^{(e, h)}$   &  1--4$^{(e)}$  \\
                        &10    & 0.9-1 AU (C)  & 1 month  & 25 \% & 2$\times$10$^{13}$--4.5$\times$10$^{13}$$^{(e, h)}$     &   11--25$^{(e)}$  \\
 \hline
\end{tabular}
\label{param_mass_dust_systems}
\begin{flushleft}
$^{(a)}$ Time spent by comets in the area defined in stellar system (column I).\\
$^{(b)}$ Column III gives the mass fraction released by comets per perihelion passage (see Fig.~\ref{fig:ratio}) in the corresponding area (column I).\\
$^{(c)}$ The mass rate released is calculated by using Eq.~(\ref{eq:fit:sigma}) multiplied 1) by the surface of a spherical comet (4$\pi$ R$_n$$^2$) and 2) by the fraction of mass released in the corresponding area (column III). The result is therefore divided by the time spent by the comets in the area (column II) to produce mass of dust produced per month per comet.\\
$^{(d)}$ The number of comets (N$_n$) is given by dividing the mass production rate (kg/year) of dust in stellar systems (see Tab.\ref{param_systems}) by 12 (months) and by the rate of mass produced by comets per month (column IV). \\
$^{(e)}$ The results are similar for a 1 km (resp. 100 km) type comet by dividing released mass rate (column IV) by 100 (resp. 10$^{-2}$), and therefore multiplying Number of comets (column V) by 100 (resp. 10$^{-2}$). \\
$^{(f)}$ Average value for comets in the solar system (Fern\'andez et al. 2013; Weiler et al. 2011; Snodgrass et al. 2011; Tancredi et al. 2006).\\
$^{(g)}$ The rate has been calculated in the area (column I) by using Eq.(\ref{eq:fit:dustprod}) and a 2.5 km radius comet. \\ 
$^{(h)}$ The minimum and maximum values correspond to the luminosities 28$^{+8}_{-6}$ L$_\odot$ and 57$\pm3$ L$_\odot$ respectively.\\
\end{flushleft}
\end{table*}

\subsection{The zodiacal dust disk}

We consider the zodiacal dust in the Solar System as a first,
  crude example of application for our model. The zodiacal dust disk
  is predominantly located between 2 AU and 3 AU (Beichman et
  al. 2005; Backman 1998). The current dust production rate in this
  inner part of the solar system is of the order of 10$^4$~kg s$^{-1}$
  (Moro-Martin 2013), 85\% of which would originate from comets
  (Nesvorny et al. 2010), with less than 10\% produced by Oort Cloud
  long-period comets (Moro-Martin 2013). By using
  Eqs. \ref{eq:fit:dustprod}, \ref{eq:fit:dustprodQ} and
  \ref{eq:fit:dustprodr}, we estimate that a typical comet of about
  2.5 km radius\footnote{JFCs have radii around 2-3 km (see
    Fern\'andez et al. 2013; Weiler et al. 2011; Snodgrass et
    al. 2011; Tancredi et al. 2006)} will release dust at an average
  rate of about 675 kg s$^{-1}$ between 2 AU and 3 AU (by using Eq.~\ref{eq:fit:dustprod} and a 2.5 km radius comet).  Considering
  only this inner part of the solar system, it thus takes at least 13
  comets in permanence in this area to maintain a dust production rate
  of 8.5$\times$10$^3$ kg s$^{-1}$ (see Tables \ref{param_systems} and \ref{param_mass_dust_systems} for the method with a 2.5 km radius comet).
This amounts to approximately 3\% of the average number of
  JFCs identified in the inner part of the solar system.

This fraction has to be compared with the fraction of the time
  spent by comets in the 2-3 AU area around the Sun. Considering an
  average period of revolution of 8 years for JFCs\footnote{This
    average period has been calculated by taking into account about
    110 periods of revolution of JFCs.}, the time spent by each comet
  in the 2-3 AU area represents approximately 4.5\% of the total time
  required to do a revolution. Assuming that comets are randomly
  distributed on their orbits around the Sun, this would mean that, on
  average, 4.5\% of the JFCs could be present at any time in the
  2-3~AU area. We remember that our model assumes no dust mantle,
  maximizing the mass production rates.  Therefore, the number of
  comets given above (13 comets) is most likely a lower
  limit. However, we note that the fraction of comets potentially
  present in the 2-3 AU area is of the same order of magnitude as the
  fraction of JFCs needed to have a dust production rate representing
  85\% of the total mass production rate ($\sim$10$^4$ kg s$^{-1}$).

\subsection{The example of Vega}
Vega, whose luminosity varies between 28$^{+8}_{-6}$ L$_\odot$ (equator) and 57$\pm3$ L$_\odot$ (poles, see Aufdenberg et al. 2006), is the archetypal debris disk star. The best model for the hot exozodiacal dust disk around the A-type star Vega suggests that it is predominantly located within 0.1--0.5\,AU of the star and has a mass of about 10$^{-9}$\,M$_{\oplus}$ (Defr\`ere et al. 2011).
An outer debris disk has also been regularly observed (Aumann et al 1984), and Sibthorpe et al. (2010) found a radius of ~85\,AU. There is no direct evidence to suggest a link between the two dusty regions, nor is there any evidence for the presence of planets orbiting Vega (Heinze et al. 2008). However if we take the Solar System as an example, we can hypothesise the presence of a population of comets with small pericentres, potentially analogous to Jupiter family or sun-grazing comets.  The models presented here can then be used to estimate the dust or gas released by these comets.

Clearly, the exact orbits and size distribution of any comets is unknown.  By considering orbits A and C, where it is the comet's pericentre that is critical, and considering both extreme stellar luminosities (28$^{+8}_{-6}$ L$_\odot$ and 57$\pm3$ L$_\odot$), a typical cometary radius of 1-10\,km can be used to determine that the total dust mass released by icy bodies during one orbit varies between (by using Eq.~\ref{eq:fit:sigma})
3$\times$10$^{12}$ -- 6$\times$10$^{14}$\,kg, and 8$\times$10$^{11}$ -- 2$\times$10$^{14}$\,kg, respectively\footnote{The minimum and maximum values correspond to the luminosities 28$^{+8}_{-6}$ L$_\odot$ and 57$\pm3$ L$_\odot$ respectively for 1 km and 10 km radius comets.}. 
Considering only the total mass released in an area of 1 AU radius around the star and per orbit (about 85\% and 25\% of the total mass for orbits A and C, respectively, see Fig.~\ref{fig:ratio}), we obtain ejected dust masses varying in the ranges 2.5$\times$10$^{12}$ -- 5$\times$10$^{14}$\,kg (4.2$\times$10$^{-13}$ -- 8.4$\times$10$^{-11}$\,M$_{\oplus}$), and 2$\times$10$^{11}$ -- 5$\times$10$^{13}$\,kg (3.3$\times$10$^{-14}$  -- 8.4$\times$10$^{-12}$\,M$_{\oplus}$), respectively.

A rough comparison of these ejected dust masses to the hot exozodiacal dust mass around Vega (about 10$^{-9}$\,M$_{\oplus}$, Defr\`ere et al. 2011), suggests that it takes between 10 and 10$^4$ passages of comets of 1-10 km radius with perihelion inferior to 1 AU, to release the observed dust. In other words, in order for the observed exozodiacal dust emission to be accounted for by dust released by comets, 10-10$^4$ comets are required to produce dust mass inside 1~AU during the dust's lifetime.
Given the high luminosity of Vega, the small estimated size of the dust grains in the observed exozodiacal, and the short orbital timescales, the dust lifetime is likely to be dominated by collisions. Absil et al. (2006) estimate a collisional lifetime of a year. Thus, using the previous estimate this corresponds to 1-2500 1--10 km radius comets passing every month (see Tables \ref{param_systems} and \ref{param_mass_dust_systems} for the method with a 10 km radius comet). 
This number should be compared to the 2200 known, essentially long-period (P$>$200\,yr) comets in the solar system (Fern{\'a}ndez 2008) and to the frequency of about one SOHO sungrazer comet every 3 days \citep[e.g.][]{bem07,sek13},  equivalent to $\sim$ 10 comets a month for solar system. 

This comparison indicates that a regular flow of typical size comets with small pericentre (i.e., of type similar to orbit A) such as to in the Solar System, could release sufficient dust to produce the exozodiacal disk observed around Vega. However, if the population of comets has larger pericentres (orbit C), a much larger population is required. Without a detailed knowledge of the planetary system orbiting Vega, it is difficult to make exact conclusions regarding what is realistic. It is, however, worth pointing out here that comets with small pericentres have much shorter lifetimes against sublimation in the Vega planetary system than our Solar System (see Sect.~\ref {lifetimecomets}), with lifetimes varying between several tens of hundreds of years to several tens of thousands of years for 1-10\,km radius type bodies, or more than 10$^5$ years for 100\,km-sized objects. 
Such a steady dusty system with short survival lifetimes for the exocomets cannot survive without a population of comets renewed from a large and cold reservoir of cometary bodies beyond the water ice line position of the stellar system, and maybe gravitationally disturbed by larger bodies such as planets (Bonsor et al. 2014; Raymond \& Bonsor 2014; Bonsor et al. 2012).

We note however that the presence of a dust mantle of several cm of thickness at the surface of exocomets could also help these bodies to survive longer by decreasing the rate of water gas production (see Marboeuf \& Schmitt 2014). In the few observations of the surface of comets in the solar system, they are found to be mainly covered by a dust mantle with small active nucleus areas (Sunshine et al. 2006, 2011; Meech et al. 2011; Thomas et al. 2015). In this case, one can expect a lower outgassing and dust ejection from the surface of comets\footnote{The formation of a dust mantle could reduce by up to 2 orders of magnitude the dust mass production at the surface of a comet (see Fig.\ref{fig:ratedust_sun})}, and smaller grains\footnote{Bigger grains stay on the surface to form the crust dust mantle.}. However, it would increase significantly by several orders of magnitude the number of typical size comets needed to reproduce the dust mass around Vega, and would confirm the transient nature of the exozodisk. Nevertheless, one or two giant comets of 100 km radius on orbit A passing once a year (lifetime of grains) and covered by a dust mantle (reducing therefore the production of dust by up to 2 orders of magnitude) could explain the observations.

We note that the cometary activity close to Vega should also
  be a source of gas, but the high radiation pressure induced by the
  high luminosity may expel the gas outward from this area on a
  shorter timescale than the dust and may leave the dust grains in an
  essentially gas-free environment close to the star.

\subsection{The example of HD 69830}
The nearby K0 V, low luminosity star HD~69830 (0.45\,L$_\odot$)
  exhibits a mass of warm dust grains of about
  4.6$\times$10$^{-9}$\,M$_{\oplus}$ (Beichman et al. 2005) within 1\,AU around the star, considering only 0.25\,$\mu$m-sized
  dust grains\footnote{We note that the assumption on the size distribution of dust grains taken in our model does not correspond to the size of grains considered around HD 69830. However, since all dust grains are ejected in space from the surface in our study, the mass of dust released by comets remains similar whatever the dust grain distribution assumed}. Such a system removes the grains by collisions,
  radiation pressure and/or Poynting-Robertson on timescales lower
  than 400--700\,years (Beichman et al. 2005), significantly less than
  the lifetime of cometary bodies in this system
  (10$^4$--10$^6$\,years, considering comets of 1--10\,km radius). The
  total mass released by bodies in an area of 1\,AU on orbits A and C
  (respectively about 100\% and 70\% of the total mass, see
  Fig.~\ref{fig:ratio}) around the star and during one revolution
  varies from about 6$\times$10$^{9}$\,kg (10$^{-15}$\,M$_{\oplus}$,
  orbit C) up to 4$\times$10$^{12}$\,kg
  (7$\times$10$^{-13}$\,M$_{\oplus}$, orbit A). Comparing these values
  to the total mass of dust grains around the star (about
  4.6$\times$10$^{-9}$\,M$_{\oplus}$, Beichman et al. 2005), it takes
  between about 7000 and 5$\times$10$^{6}$ comets for orbits A and C,
  respectively, per unit of dust removing timescale. Assuming a
  reference lifetime for the dust of 400\,years (see Beichman 2005),
it needs between 2 (orbit A) and 1000 (orbit C) 1--10\,km radius
comets per month to supply the disk in dust grains, depending on the
location of the grains around the star, and perihelion position of the
cometary bodies (see Tables \ref{param_systems} and \ref{param_mass_dust_systems} for the method with a 10 km radius comet). By increasing the radius of objects to 100\,km, the flow of comets varies from 1 giant comet every 6 years to 10 giant comets per month.
By considering a full grain size distribution like for the Vega
models, \citet{olo12} showed that the debris disk dust mass could
increase up to a few 10$^{-7}$\,M$_{\oplus}$. In this case, one can
expect a higher (by two orders of magnitude) and perhaps unrealistic
number of comets transiting close to the star per month.

Contrary to Vega, the lifetime of the comets around HD~69830
  is much less of a problem since they are removed after several tens
  of thousands up to a few millions of years, depending on their
  radius. However, would the HD~69830 dusty system be in steady state,
  it would require a sufficiently high number of comets to sustain the
  debris disk for more than 3\,Gyr, which requires a large reservoir
  of comets in colder regions of the system, and, like for Vega,
  gravitational perturbers to generate comets. However, the
  non-detection of cold dust with Herschel by \citet{mar14} does not
  provide strong support to such a scenario. Instead, the warm
  exozodiacal dust disk around HD~69830 could most likely be a
  transient phenomenon occurring only on several hundreds of years as
  proposed earlier by Beichman et al. (2005) and \citet{olo12} for
  instance.

\section{Discussion \& conclusion \label{discussion}}

The thermal evolution and physical alteration of comets in stellar systems has been studied for several stellar luminosities and orbital parameters. To do this, we have used a quasi 3D model of cometary nucleus representing a spherical comet composed of dust grains and H$_2$O ice.
Results provide mass productions of dust and gas H$_2$O and total mass of cometary material ejected per revolution for stellar luminosities varying from 0.1 to 70 $L_\odot$ and distances to the star varying from 0.1 to 50 AU. The dust/gas mass ratio ejected by comets is shown to be approximately of the same order of magnitude than the dust/ice mass ratio in the nucleus. In order to use these results by any model, and to compare observations with outgassing and dust released by comets, we also provided simple laws allowing to reproduce easily these data for one comet as a function of stellar luminosities and distance to the star. In summary, we have provided laws for average thickness of cometary material removed, total mass lost per revolution by comets, lifetime of these objects as mass production of H$_2$O molecules and dust grains as a function of stellar luminosities, orbits and distance to the stars. \\

We show that the physical alteration suffered by comets changes by some order of magnitude as a function of the stellar luminosity $L_{\star}$ and orbit. High temperatures reached at the surface of bodies for high stellar luminosities induce high average thickness of cometary material (up to $\approx$ 2 km) removed from the surface per revolution. For high stellar luminosity (greater than 10 $L_\odot$), the temperature reached at the surface and the resulting thickness of cometary material removed is so high that comets behave like new bodies exposing layers never physically altered in surface after perihelion passage. For stellar luminosity of 70 $L_\odot$, the total thickness of cometary material removed at equator exceeds by 1 km the one at the poles leading to a nucleus eaten in the center, thereby changing greatly the original spherical shape of the comet. 
Such thicknesses of cometary material removed per revolution
exceeds by far the physical alteration encountered by comets in
solar system (of the order of magnitude of the meter). Moreover, the
difference of radius between day side and night side of the nucleus
can reach 100 m at perihelion and equator.
As a result, it is not certain that cometary nucleus
is not deviated from its initial orbit since the center of gravity of
the body changes slightly with time and mass ejection from the day
side reaches 100 times the solar system one. In addition, comets could not
survive to the important physical alteration suffered by nuclei
because the cohesion of cometary material between the poles and
equator could be broken.

We have shown that the total mass of cometary material lost in vacuum
space is mainly function of the size of objects, their orbital
parameters and the stellar luminosity $L_{\star}$ but remains approximately
constant whatever the porosity and thermal conductivity of the
cometary material: only the thickness of cometary material removed
changes. 
The total mass lost per unit of cometary surface area and per
revolution follows a power law with $L_{\star}$ and varies of several orders
of magnitude from 10 to 10$^6$ kg m$^{-2}$ for stellar luminosities
varying from 0.1 to 70 $L_\odot$, whatever the orbital parameters used
in the study. We have shown that the calculated lifetime $\tau$ of
comets changes by some order of magnitude as a function of their
initial radius $R_n$, stellar luminosity $L_{\star}$ and orbital
parameters. The lifetime of objects varies linearly with $R_n$ and
follows a power law with $L_{\star}$. The lifetime of comets, in addition to the production rates of water gas and dust, can help to disentangle scenarios assuming steady state cometary dust production from transient events (e.g. catastrophic collisions) to explain warm and hot exozodiacal dust disks around main sequence stars.
In our study, the lifetime of comets of 10 km radius varies from 6$\times$10$^4$ to 6$\times$10$^5$ years in solar system. It decreases to some thousands of years or less for stellar luminosities greater than 28 $L_\odot$. The lifetime of comets smaller than or equal to 1 km radius can not exceed 500 years whatever the orbital parameters used in the study. As discussed in Sect.~\ref{lifetimecomets} and \ref{disks}, by studying the debris disks around Vega, we have shown that a regular flow of typical size comets with perihelion close to the star (orbit A) such as in the solar system could explain the observations of the hot exozodiacal disk of Vega. However, since the lifetime of these small bodies is short in regard of the solar system, such massive debris disk cannot survive without considering a large and cold reservoir of cometary bodies beyond the water ice line position of the stellar system, and gravitationally disturbed by planets\footnote{Distant perturbations by planets such as resonances could excite icy bodies onto planet-crossing orbits (see Dones et al. 2004).} (Bonsor et al. 2014; Raymond \& Bonsor 2014; Bonsor et al. 2012) such as in the solar system (Jewitt 2004; Duncan et al. 2004) to supply the inner region in comets, i.e. dust grains. Without planets placed beyond the snow line and disturbing reservoir of icy bodies, the comet assumption does not remain valid. We note however that a giant comet of 100 km radius on orbit A, with a lifetime greater than 10$^5$ years, and passing once a year could also explain the observations.
Inversely, although the stellar luminosity of HD~69830 is low and comet type bodies are removed only after several thousand of years, we have shown by considering a full size distribution of dust grains (Olofsson et al. 2012) that the massive debris disk close to the star can be explained only by considering an unrealistic number of typical size comets transiting close to the star in the best scenario. Finally, the massive dust disk around HD~69830 could most be a transient phenomenon as proposed earlier by Beichman et al. (2005) and \citet{olo12}.

We note that the general laws provided in this study, and conclusions about the pertinence of the presence of comets around stellar systems have
required some assumptions and simplifications on the cometary nucleus model. First, we have adopted only the most abundant
chemical species H$_2$O in comets. If exocomets are depleted in H$_2$O molecules and are mainly made of
carbonaceous elements such as CO, CO$_2$ and CH$_3$OH, which sublimate at lower temperature (see Marboeuf et al. 2008), the thermodynamic
behavior, physical alteration (lifetime), and the gas and dust mass productions of comets would be considerably changed compared to the present study. 
Second, the mass of dust and H$_2$O produced in this study, given as a function of the stellar luminosity, are the maximum
productions leaving a comet since the model does not consider the formation of a dust mantle (all is ejected in space) at the surface of the comet.
The formation of such dust mantle as observed at the surface of solar system comets would decrease the mass production rate of water and dust (see Marboeuf \& Schmitt 2014), and consequently increase their lifetime. Unfortunately, it also increases significantly the number of objects needed to supply the inner regions in dust grains.
Moreover, the orbital changes related to gravitational
interactions with planets are not taken into account although it
could change significantly the orbital parameters and hence the
thermodynamic and physico-chemical behavior of comets.
In addition, we have adopted a spherical nucleus with
obliquity of 0 degree for the comet. By taking into account a
obliquity of 90 degree, the same surface could be exposed to the star
during a long period (without diurnal variations) and could change both the shape of comets and their dust and gas mass productions. 
In addition, if comets have a dust/ice mass ratio greatly different from 1, the density, the mass and the thermodynamic behavior of comets could change significantly compared to this study: a higher dust/ice mass ratio would increase the rate of dust grains ejected although a lower value should decrease it. Finally, whatever the dust and gas mass productions, we have always considered that the surface of comets were fully exposed to the star (except the night side), assuming the sublimation of water ice into free space. Yet, a high outgassing and dust ejection from the comet near pericenter could 1) partially hide the surface of the nucleus from the star and decrease substantially the stellar luminosity, and 2) create a important gas phase around comets invalidating the hypothesis of vacuum space around the nucleus. As a result, the
dust and gas mass productions of comets near the star should decrease, especially for high stellar luminosity.
Finally, our predictions of water outgassing and dust released by
comets have to be taken with caution. They probably represent the
maximum rate productions around stars. Consequently, the number of
comets and their lifetime could be higher than the estimations given in the study.

The authors thank Philippe Thebault that helped us to improve substantially the presentation of our results.
We thank the French National Research Agency (ANR) for financial support through contract ANR-2010 BLAN-0505-01 (EXOZODI). U.M. also thanks the French Space Agency, Centre National d'Etudes Spatiales (CNES), and the Swiss National Science Foundation and the Center for Space and Habitability of the University of Bern. All the computations presented in this paper were performed at the Service Commun de Calcul Intensif de l'Observatoire de Grenoble (SCCI).


\begin{thebibliography}{00}

\bibitem[Absil et al.(2013)]{2013A&A...555A.104A} Absil, O., Defr{\`e}re, D., Coud{\'e} du Foresto, V., et al.\ 2013, A\&A, 555, AA104 

\bibitem[Absil et al.(2006)]{2006SPIE.6268E...9A} Absil, O., Di Folco, E., M{\'e}rand, A., et al.\ 2006, spie, 6268

\bibitem[A'Hearn et al.(2005)]{2005Sci...310..258A} A'Hearn, M.~F., Belton, M.~J.~S., Delamere, W.~A., et al.\ 2005, Science, 310, 258 

\bibitem[Alcock et al.(1986)]{1986ApJ...302..462A} Alcock, C., Fristrom, C.~C., \& Siegelman, R.\ 1986, ApJ, 302, 462

\bibitem[Aufdenberg et al.(2006)]{2006ApJ...645..664A} Aufdenberg, J.~P., M{\'e}rand, A., Coud{\'e} du Foresto, V., et al.\ 2006, ApJ, 645, 664


\bibitem[Backman et al.(1998)]{1998Ap&SS.255...91B} Backman, D.~E., Fajardo-Acosta, S.~B., Stencel, R.~E., \& Stauffer, J.~R.\ 1998, Ap\&SS, 255, 91 

%

\bibitem[Beichman et al.(2005)]{2005ApJ...626.1061B} Beichman, C.~A., Bryden, G., Gautier, T.~N., et al.\ 2005, ApJ, 626, 1061


\bibitem[Bemporad et al.(2007)]{bem07} Bemporad, A., Poletto, G., Raymond, J., \& Giordano, S.\ 2007, Planetary Space Science, 55, 1021


\bibitem[Beust et al.(2001)]{2001A&A...366..945B} Beust, H., Karmann, C., \& Lagrange, A.-M.\ 2001, A\&A, 366, 945 

\bibitem[Beust \& Morbidelli(2000)]{2000Icar..143..170B} Beust, H., \& Morbidelli, A.\ 2000, Icarus, 143, 170 

\bibitem[Beust \& Morbidelli(1996)]{1996Icar..120..358B} Beust, H., \& Morbidelli, A.\ 1996, Icarus, 120, 358 

\bibitem[Beust \& Lissauer(1994)]{1994A&A...282..804B} Beust, H., \& Lissauer, J.~J.\ 1994, A\&A, 282, 804

\bibitem[Beust et al.(1991)]{1991A&A...247..505B} Beust, H., Vidal-Madjar, A., \& Ferlet, R.\ 1991, A\&A, 247, 505 

\bibitem[Beust et al.(1990)]{1990A&A...236..202B} Beust, H., Vidal-Madjar, A., Ferlet, R., \& Lagrange-Henri, A.~M.\ 1990, A\&A, 236, 202 

\bibitem[Beust et al.(1989)]{1989ESASP.302..167B} Beust, H., Lagrange-Henri, A.~M., Vidal-Madjar, A., \& Ferlet, R.\ 1989, Physics and Mechanics of Cometary Materials, 302, 167

\bibitem[Bonsor et al.(2014)]{2014MNRAS.441.2380B} Bonsor, A., Raymond, S.~N., Augereau, J.-C., \& Ormel, C.~W.\ 2014, MNRAS, 441, 2380

\bibitem[Bonsor et al.(2012)]{2012A&A...548A.104B} Bonsor, A., Augereau, J.-C., \& Th{\'e}bault, P.\ 2012, A\&A, 548, A104 

\bibitem[Bonsor et al.(2011)]{2011MNRAS.414..930B} Bonsor, A., Mustill, A.~J., \& Wyatt, M.~C.\ 2011, MNRAS, 414, 930 

\bibitem[Boogert \& Ehrenfreund(2004)]{2004ASPC..309..547B} Boogert, A.~C.~A., \& Ehrenfreund, P.\ 2004, Astrophysics of Dust, 309, 547 

\bibitem[Bockel{\'e}e-Morvan et al.(2009)]{2009A&A...505..825B} Bockel{\'e}e-Morvan, D., Henry, F., Biver, N., et al.\ 2009, A\&A, 505, 825

\bibitem[Bockel{\'e}e-Morvan et al.(2004)]{2004come.book..391B} Bockel{\'e}e-Morvan, D., Crovisier, J., Mumma, M.~J., \& Weaver, H.~A.\ 2004, Comets II, 391

\bibitem[Burns et al.(1979)]{1979Icar...40....1B} Burns, J.~A., Lamy, P.~L., \& Soter, S.\ 1979, Icarus, 40, 1

\bibitem[Carman(1956)]{} Carman, P. C.\ 1956, London: Butter worths Scientific Publications

\bibitem[Cohen et al.(2003)]{2003NewA....8..179C} Cohen, M., Prialnik, D., \& Podolak, M.\ 2003, New A, 8, 179 


\bibitem[Cox \& Pilachowski(2000)]{2000PhT....53j..77C} Cox, A.~N., \& Pilachowski, C.~A.\ 2000, Physics Today, 53, 77

\bibitem[Colom et al.(1997)]{1997EM&P...78...37C} Colom, P., G{\'e}rard, E., Crovisier, J., et al.\ 1997, Earth Moon and Planets, 78, 37 

\bibitem[Combi et al.(2013)]{2013Icar..225..740C} Combi, M.~R., M{\"a}kinen, J.~T.~T., Bertaux, J.-L., et al.\ 2013, Icarus, 225, 740 

\bibitem[Combi et al.(2011)]{2011Icar..216..449C} Combi, M.~R., Boyd, Z., Lee, Y., et al.\ 2011a, Icarus, 216, 449 

\bibitem[Combi et al.(2011)]{2011ApJ...734L...6C} Combi, M.~R., Bertaux, J.-L., Qu{\'e}merais, E., Ferron, S., Makinen, J.~T.~T.\ 2011b, ApJl, 734, L6

\bibitem[Combi et al.(2011)]{2011AJ....141..128C} Combi, M.~R., Lee, Y., Patel, T.~S., et al.\ 2011c, AJ, 141, 128

\bibitem[Combi et al.(2009)]{2009AJ....137.4734C} Combi, M.~R., M{\"a}kinen, J.~T.~T., Bertaux, J.-L., Lee, Y., \& Qu{\'e}merais, E.\ 2009, AJ, 137, 4734 

\bibitem[Crifo et al.(1997)]{1997A&A...320L..29C} Crifo, F., Vidal-Madjar, A., Lallement, R., Ferlet, R., \& Gerbaldi, M.\ 1997, A\&A, 320, L29 

\bibitem[Dartois(2009)]{2009ASPC..414..411D} Dartois, E.\ 2009, Cosmic Dust - Near and Far, 414, 411

\bibitem[Dauphas(2003)]{2003Icar..165..326D} Dauphas, N.\ 2003, Icarus, 165, 326 

\bibitem[Davidsson et al.(2013)]{2013Icar..224..154D} Davidsson, B.~J.~R., Guti{\'e}rrez, P.~J., Groussin, O., et al.\ 2013, Icarus, 224, 154 

\bibitem[Davidsson et al.(2009)]{2009Icar..201..335D} Davidsson, B.~J.~R., Guti{\'e}rrez, P.~J., \& Rickman, H.\ 2009, Icarus, 201, 335

\bibitem[Davidsson et al.(2007)]{2007Icar..187..306D} Davidsson, B.~J.~R., Guti{\'e}rrez, P.~J., \& Rickman, H.\ 2007, Icarus, 187, 306

\bibitem[Davidsson \& Guti{\'e}rrez(2006)]{2006Icar..180..224D} Davidsson, B.~J.~R., \& Guti{\'e}rrez, P.~J.\ 2006, Icarus, 180, 224

\bibitem[Davidsson \& Guti{\'e}rrez(2005)]{2005Icar..176..453D} Davidsson, B.~J.~R., \& Guti{\'e}rrez, P.~J.\ 2005, Icarus, 176, 453

\bibitem[Davidsson \& Guti{\'e}rrez(2004)]{2004Icar..168..392D} Davidsson, B.~J.~R., \& Guti{\'e}rrez, P.~J.\ 2004, Icarus, 168, 392

\bibitem[Davidsson \& Skorov(2002)]{2002Icar..159..239D} Davidsson, B.~J.~R., \& Skorov, Y.~V.\ 2002, Icarus, 159, 239

\bibitem[Debes et al.(2012)]{2012ApJ...754...59D} Debes, J.~H., Kilic, M., Faedi, F., et al.\ 2012, ApJ, 754, 59 

\bibitem[Debes \& Sigurdsson(2002)]{2002ApJ...572..556D} Debes, J.~H., \& Sigurdsson, S.\ 2002, ApJ, 572, 556 

\bibitem[Defr{\`e}re et al.(2011)]{2011A&A...534A...5D} Defr{\`e}re, D., Absil, O., Augereau, J.-C., et al.\ 2011, A\&A, 534, AA5

\bibitem[Dello Russo et al.(2002)]{2002JGRE..107.5095D} Dello Russo, N., Mumma, M.~J., DiSanti, M.~A., \& Magee-Sauer, K.\ 2002, Journal of Geophysical Research (Planets), 107, 5095

\bibitem[Delsemme \& Miller(1971)]{1971P&SS...19.1229D} Delsemme, A.~H., \& Miller, D.~C.\ 1971, Planetary Space Science, 19, 1229 

\bibitem[de Vries et al.(2012)]{2012Natur.490...74D} de Vries, B.~L., Acke, B., Blommaert, J.~A.~D.~L., et al.\ 2012, Nature, 490, 74

\bibitem[Dones et al.(2004)]{2004come.book..153D} Dones, L., Weissman, P.~R., Levison, H.~F., \& Duncan, M.~J.\ 2004, Comets II, 153

\bibitem[Duncan et al.(2004)]{2004come.book..193D} Duncan, M., Levison, H., \& Dones, L.\ 2004, Comets II, 193 

\bibitem[Eiroa et al.(2013)]{eiroa2013} Eiroa, C., Marshall, J.~P., Mora, A., et al.\ 2013, A\&A, 555, A11 

\bibitem[Ertel et al.(2014)]{2014A&A...570A.128E} Ertel, S., Absil, O., Defr{\`e}re, D., et al.\ 2014, A\&A, 570, AA128 

\bibitem[Espinasse et al.(1991)]{1991Icar...92..350E} Espinasse, S., Klinger, J., Ritz, C., \& Schmitt, B.\ 1991, Icarus, 92, 350 

\bibitem[Fanale \& Salvail(1984)]{1984Icar...60..476F} Fanale, F.~P., \& Salvail, J.~R.\ 1984, Icarus, 60, 476


\bibitem[Farihi et al.(2009)]{2009ApJ...694..805F} Farihi, J., Jura, M., \& Zuckerman, B.\ 2009, ApJ, 694, 805 

\bibitem[Feistel \& Wagner(2006)]{2006JPCRD..35.1021F} Feistel, R., \& Wagner, W.\ 2006, Journal of Physical and Chemical Reference Data, 35, 1021

\bibitem[Fern{\'a}ndez(2002)]{2002EM&P...89....3F} Fern{\'a}ndez, Y.~R.\ 2002, Earth Moon and Planets, 89, 3 

\bibitem[Fern{\'a}ndez(2008)]{2008SSRv..138...27F} Fern{\'a}ndez, J.~A.\ 2008,  Space Science Rev., 138, 27 


\bibitem[Fern{\'a}ndez et al.(2013)]{2013Icar..226.1138F} Fern{\'a}ndez, Y.~R., Kelley, M.~S., Lamy, P.~L., et al.\ 2013, Icarus, 226, 1138


\bibitem[Ford \& Neufeld(2001)]{2001ApJ...557L.113S} Saavik Ford, K.~E.~S., \& Neufeld, D.~A.\ 2001, ApJl, 557, L113 

\bibitem[Ford et al.(2003)]{2003ApJ...589..430F} Ford, K.~E.~S., Neufeld, D.~A., Goldsmith, P.~F., \& Melnick, G.~J.\ 2003, ApJ, 589, 430 

\bibitem[Ford et al.(2004)]{2004ApJ...614..990F} Ford, K.~E.~S., Neufeld, D.~A., Schilke, P., \& Melnick, G.~J.\ 2004, ApJ, 614, 990 

\bibitem[Fulle et al.(1997)]{1997A&A...324.1197F} Fulle, M., Mikuz, H., \& Bosio, S.\ 1997, A\&A, 324, 1197 

\bibitem[Fray \& Schmitt(2009)]{2009P&SS...57.2053F} Fray, N., \& Schmitt, B.\ 2009,  Planetary Space Science, 57, 2053

\bibitem[G{\"a}nsicke et al.(2006)]{2006Sci...314.1908G} G{\"a}nsicke, B.~T., Marsh, T.~R., Southworth, J., \& Rebassa-Mansergas, A.\ 2006, Science, 314, 1908 

\bibitem[Gibb et al.(2000)]{2000ApJ...536..347G} Gibb, E.~L., Whittet, D.~C.~B., Schutte, W.~A., et al.\ 2000, ApJ, 536, 347 

\bibitem[Girven et al.(2012)]{2012ApJ...749..154G} Girven, J., Brinkworth, C.~S., Farihi, J., et al.\ 2012, ApJ, 749, 154 

\bibitem[Gortsas et al.(2011)]{2011Icar..212..858G} Gortsas, N., K{\"u}hrt, E., Motschmann, U., \& Keller, H.~U.\ 2011, Icarus, 212, 858 

\bibitem[Grady et al.(1996)]{1996ApJ...471L..49G} Grady, C.~A., Perez, M.~R., Talavera, A., et al.\ 1996, ApJl, 471, L49 

\bibitem[Greenberg(1982)]{1982come.coll..131G} Greenberg, J.~M.\ 1982, IAU Colloq.~61: Comet Discoveries, Statistics, and Observational Selection, 131


\bibitem[Guti{\'e}rrez et al.(2000)]{2000A&A...355..809G} Guti{\'e}rrez, P.~J., Ortiz, J.~L., Rodrigo, R., \& L{\'o}pez-Moreno, J.~J.\ 2000, A\&A, 355, 809 

\bibitem[Hainaut(2011)]{2011ASPC..450..233H} Hainaut, O.~R.\ 2011, Astronomical Society of the Pacific Conference Series, 450, 233 

\bibitem[Heinze et al.(2008)]{2008ApJ...688..583H} Heinze, A.~N., Hinz, P.~M., Kenworthy, M., Miller, D., \& Sivanandam, S.\ 2008, ApJ, 688, 583

\bibitem[Horanyi et al.(1984)]{1984ApJ...278..449H} Horanyi, M., Gombosi, T.~I., Cravens, T.~E., et al.\ 1984, ApJ, 278, 449 


\bibitem[Huebner et al.(2006)]{2006hgdc.conf.....H} Huebner, W.~F., Benkhoff, J., Capria, M.-T., Coradini, A., de Sanctis, C., Orosei, R., 
\& Prialnik, D.\ 2006, Published for The International Space Science Institute, Bern, Switzerland, by ESA Publications Division, Noordwijk, The Netherlands.

\bibitem[Irvine et al.(2000)]{2000prpl.conf.1159I} Irvine, W.~M., Schloerb, F.~P., Crovisier, J., Fegley, B., Jr., \& Mumma, M.~J.\ 2000, Protostars and Planets IV, 1159

\bibitem[Jewitt(2004)]{2004come.book..659J} Jewitt, D.~C.\ 2004, Comets II, 659

\bibitem[Jewitt \& Matthews(1999)]{1999AJ....117.1056J} Jewitt, D., \& Matthews, H.\ 1999, AJ, 117, 1056

\bibitem[Jura(2008)]{2008AJ....135.1785J} Jura, M.\ 2008, AJ, 135, 1785 

\bibitem[Jura(2005)]{2005AJ....130.1261J} Jura, M.\ 2005a, AJ, 130, 1261 

\bibitem[Jura(2005)]{2005ApJ...620..487J} Jura, M.\ 2005b, ApJ, 620, 487 

\bibitem[Julian et al.(2000)]{2000Icar..144..160J} Julian, W.~H., Samarasinha, N.~H., \& Belton, M.~J.~S.\ 2000, Icarus, 144, 160 

\bibitem[Karmann et al.(2003)]{2003A&A...409..347K} Karmann, C., Beust, H., \& Klinger, J.\ 2003, A\&A, 409, 347 

\bibitem[Karmann et al.(2001)]{2001A&A...372..616K} Karmann, C., Beust, H., \& Klinger, J.\ 2001, A\&A, 372, 616 

\bibitem[Kidger(2003)]{2003A&A...408..767K} Kidger, M.~R.\ 2003, A\&A, 408, 767

\bibitem[Kiefer et al.(2014)]{2014A&A...561L..10K} Kiefer, F., Lecavelier des Etangs, A., Augereau, J.-C., et al.\ 2014a, A\&A, 561, LL10

\bibitem[Kiefer et al.(2014)]{2014Natur.514..462K} Kiefer, F., Lecavelier des Etangs, A., Boissier, J., et al.\ 2014b, Nature, 514, 462

\bibitem[Kossacki et al.(1999)]{1999Icar..142..202K} Kossacki, K.~J., Szutowicz, S.~{\L}., \& Leliwa-Kopysty{\'n}ski, J.\ 1999, Icarus, 142, 202

\bibitem[Klein et al.(2010)]{2010ApJ...709..950K} Klein, B., Jura, M., Koester, D., Zuckerman, B., \& Melis, C.\ 2010, ApJ, 709, 950

\bibitem[Kofman et al.(2015)]{2015Sci...349b0639K} Kofman, W., Herique, A., Barbin, Y., et al.\ 2015, Science, 349, 020639 

\bibitem[Kossacki \& Szutowicz(2006)]{2006P&SS...54...15K} Kossacki, K.~J., \& Szutowicz, S.\ 2006, Planetary Space Science, 54, 15 

\bibitem[Lamers et al.(1997)]{1997A&A...328..321L} Lamers, H.~J.~G.~L.~M., Lecavelier Des Etangs, A., \& Vidal-Madjar, A.\ 1997, A\&A, 328, 321 

\bibitem[Lamy et al.(2004)]{2004come.book..223L} Lamy, P.~L., Toth, I., Fernandez, Y.~R., \& Weaver, H.~A.\ 2004, Comets II, 223

\bibitem[Lamy et al.(2007)]{2007SSRv..128...23L} Lamy, P.~L., Toth, I., Davidsson, B.~J.~R., et al.\ 2007, Space Science Rev., 128, 23

\bibitem[Langer et al.(2000)]{2000prpl.conf...29L} Langer, W.~D., van Dishoeck, E.~F., Bergin, E.~A., et al.\ 2000, Protostars and Planets IV, 29

\bibitem[Lara et al.(2004)]{2004A&A...420..371L} Lara, L.-M., Rodrigo, R., Tozzi, G.~P., Boehnhardt, H., \& Leisy, P.\ 2004, A\&A, 420, 371 

\bibitem[Lasue et al.(2008)]{2008P&SS...56.1977L} Lasue, J., de Sanctis, M.~C., Coradini, A., et al.\ 2008, Planetary Space Science, 56, 1977 

\bibitem[Lebreton et al.(2012)]{2012A&A...539A..17L} Lebreton, J., Augereau, J.-C., Thi, W.-F., et al.\ 2012, A\&A, 539, A17

\bibitem[Lecavelier Des Etangs(1999)]{1999A&AS..140...15L} Lecavelier Des Etangs, A.\ 1999a, A\&As, 140, 15 

\bibitem[Lecavelier Des Etangs et al.(1999)]{1999A&A...343..916L} Lecavelier Des Etangs, A., Vidal-Madjar, A., \& Ferlet, R.\ 1999b, A\&A, 343, 916 

\bibitem[Lecavelier Des Etangs et al.(1996)]{1996A&A...307..542L} Lecavelier Des Etangs, A., Vidal-Madjar, A., \& Ferlet, R.\ 1996, A\&A, 307, 542

\bibitem[Levison et al.(2001)]{2001AJ....121.2253L} Levison, H.~F., Dones, L., \& Duncan, M.~J.\ 2001, AJ, 121, 2253

\bibitem[Levison \& Duncan(1997)]{1997Icar..127...13L} Levison, H.~F., \& Duncan, M.~J.\ 1997, Icarus, 127, 13

\bibitem[Li \& Greenberg(1998)]{1998A&A...331..291L} Li, A., \& Greenberg, J.~M.\ 1998, A\&A, 331, 291 

\bibitem[Lisse et al.(2009)]{2009PASP..121..968L} Lisse, C.~M., Fernandez, Y.~R., Reach, W.~T., et al.\ 2009, PASP, 121, 968 

\bibitem[Lisse et al.(2002)]{2002dsso.conf..259L} Lisse, C.~M., A'Hearn, M.~F., Fernandez, Y.~R., \& Peschke, S.~B.\ 2002, IAU Colloq.~181: Dust in the Solar System and Other Planetary Systems, 259 

\bibitem[Lodders(2003)]{2003ApJ...591.1220L} Lodders, K.\ 2003, ApJ, 591, 1220 

\bibitem[Lowry et al.(2012)]{2012A&A...548A..12L} Lowry, S., Duddy, S.~R., Rozitis, B., et al.\ 2012, A\&A, 548, A12

\bibitem[Lowry et al.(2008)]{2008ssbn.book..397L} Lowry, S., Fitzsimmons, A., Lamy, P., \& Weissman, P.\ 2008, The Solar System Beyond Neptune, 397

\bibitem[Lowry \& Weissman(2007)]{2007Icar..188..212L} Lowry, S.~C., \& Weissman, P.~R.\ 2007, Icarus, 188, 212


\bibitem[Marboeuf \& Schmitt(2014)]{2014Icar..242..225M} Marboeuf, U., \& Schmitt, B.\ 2014, Icarus, 242, 225

\bibitem[Marboeuf et al.(2014)]{2014A&A...570A..35M} Marboeuf, U., Thiabaud, A., Alibert, Y., Cabral, N., \& Benz, W.\ 2014, A\&A, 570, AA35

\bibitem[Marboeuf et al.(2012)]{2012A&A...542A..82M} Marboeuf, U., Schmitt, B., Petit, J.-M., Mousis, O., \& Fray, N.\ 2012, A\&A, 542, A82

\bibitem[Marboeuf et al.(2008)]{2008ApJ...681.1624M} Marboeuf, U., Mousis, O., Ehrenreich, D., Alibert, Y., Cassan, A., Wakelam, V., \& Beaulieu, J.-P.\ 2008, ApJ, 681, 1624 

\bibitem[Marshall et al.(2014)]{mar14} Marshall, J.~P., Moro-Mart{\'{\i}}n, A., Eiroa, C., et al.\ 2014, A\&A, 565, AA15

\bibitem[McDonnell et al.(1987)]{1987A&A...187..719M} McDonnell, J.~A.~M., et al.\ 1987, A\&A, 187, 719

\bibitem[McDonnell et al.(1986)]{1986Natur.321..338M} McDonnell, J.~A.~M., Alexander, W.~M., Burton, W.~M., et al.\ 1986, Nature, 321, 338

\bibitem[Meech et al.(2011)]{2011ApJ...734L...1M} Meech, K.~J., A'Hearn, M.~F., Adams, J.~A., et al.\ 2011, ApJl, 734, LL1

\bibitem[Mekler et al.(1990)]{1990ApJ...356..682M} Mekler, Y., Prialnik, D., \& Podolak, M.\ 1990, ApJ, 356, 682 

\bibitem[Melnick et al.(2001)]{2001Natur.412..160M} Melnick, G.~J., Neufeld, D.~A., Ford, K.~E.~S., Hollenbach, D.~J., \& Ashby, M.~L.~N.\ 2001, Nature, 412, 160

\bibitem[Montgomery \& Welsh(2012)]{2012PASP..124.1042M} Montgomery, S.~L., \& Welsh, B.~Y.\ 2012, PASP, 124, 1042

\bibitem[Moreno et al.(2012)]{2012ApJ...752..136M} Moreno, F., Pozuelos, F., Aceituno, F., et al.\ 2012, ApJ, 752, 136 

\bibitem[Moro-Martin(2013)]{2013pss3.book..431M} Moro-Martin, A.\ 2013, Planets, Stars and Stellar Systems.~Volume 3: Solar and Stellar Planetary Systems, 431

\bibitem[Mottola et al.(2014)]{2014A&A...569L...2M} Mottola, S., Lowry, S., Snodgrass, C., et al.\ 2014, A\&A, 569, LL2 

\bibitem[Mumma \& Charnley(2011)]{2011ARA&A..49..471M} Mumma, M.~J., \& Charnley, S.~B.\ 2011, ARA\&A, 49, 471 

\bibitem[Mumma(1997)]{1997ASPC..122..369M} Mumma, M.~J.\ 1997, From Stardust to Planetesimals, 122, 369

\bibitem[Nesvorn{\'y} et al.(2010)]{2010ApJ...713..816N} Nesvorn{\'y}, D., Jenniskens, P., Levison, H.~F., et al.\ 2010, ApJ, 713, 816

\bibitem[Olofsson et al.(2012)]{olo12} Olofsson, J., Juh{\'a}sz, A., Henning, T., et al.\ 2012, A\&A, 542, AA90

\bibitem[Prialnik et al.(2004)]{2004come.book..359P} Prialnik, D., Benkhoff, J., \& Podolak, M.\ 2004, Comets II, 359 

\bibitem[Prialnik(1997)]{1997ApJ...478L.107P} Prialnik, D.\ 1997, ApJl, 478, L107

\bibitem[Raymond \& Bonsor(2014)]{2014MNRAS.442L..18R} Raymond, S.~N., \& Bonsor, A.\ 2014, MNRAS, 442, L18

\bibitem[Richardson et al.(2007)]{2007Icar..190..357R} Richardson, J.~E., Melosh, H.~J., Lisse, C.~M., \& Carcich, B.\ 2007, Icarus, 190, 357

\bibitem[Rickman et al.(1990)]{1990A&A...237..524R} Rickman, H., Fernandez, J.~A., \& Gustafson, B.~A.~S.\ 1990, A\&A, 237, 524

\bibitem[Roberge et al.(2002)]{2002ApJ...568..343R} Roberge, A., Feldman, P.~D., Lecavelier des Etangs, A., et al.\ 2002, ApJ, 568, 343 

\bibitem[Rosenberg \& Prialnik(2007)]{2007NewA...12..523R} Rosenberg, E.~D., \& Prialnik, D.\ 2007, New A, 12, 523

\bibitem[Sanzovo et al.(2001)]{2001MNRAS.326..852S} Sanzovo, G.~C., de Almeida, A.~A., Misra, A., et al.\ 2001, MNRAS, 326, 852

\bibitem[Schleicher(2006)]{2006Icar..181..442S} Schleicher, D.~G.\ 2006, Icarus, 181, 442

\bibitem[Sekanina(1979)]{1979Icar...37..420S} Sekanina, Z.\ 1979, Icarus, 37, 420

\bibitem[Sekanina \& Kracht(2013)]{sek13} Sekanina, Z., \& Kracht, R.\ 2013, ApJ, 778, 24

\bibitem[Sibthorpe et al.(2010)]{2010A&A...518L.130S} Sibthorpe, B., Vandenbussche, B., Greaves, J.~S., et al.\ 2010, A\&A, 518, LL130 


\bibitem[Sierks et al. (2015)]{} Sierks, H., Barbieri, C., Lamy, P.L., et al.\ 2015, Science, 347, aaa1044

\bibitem[Snodgrass et al.(2011)]{2011MNRAS.414..458S} Snodgrass, C., Fitzsimmons, A., Lowry, S.~C., \& Weissman, P.\ 2011, MNRAS, 414, 458 

\bibitem[Sosa \& Fern{\'a}ndez(2009)]{2009MNRAS.393..192S} Sosa, A., \& Fern{\'a}ndez, J.~A.\ 2009, MNRAS, 393, 192

\bibitem[Sosa \& Fern{\'a}ndez(2011)]{2011MNRAS.416..767S} Sosa, A., \& Fern{\'a}ndez, J.~A.\ 2011, MNRAS, 416, 767 

\bibitem[Stern et al.(1990)]{1990Natur.345..305S} Stern, S.~A., Shull, J.~M., \& Brandt, J.~C.\ 1990, Nature, 345, 305 

\bibitem[Su et al.(2006)]{2006ApJ...653..675S} Su, K.~Y.~L., Rieke, G.~H., Stansberry, J.~A., et al.\ 2006, ApJ, 653, 675 

\bibitem[Sunshine et al.(2011)]{2011epsc.conf.1345S} Sunshine, J.~M., Feaga, L.~M., Groussin, O., et al.\ 2011, EPSC-DPS Joint Meeting 2011, 1345

\bibitem[Sunshine et al.(2006)]{2006Sci...311.1453S} Sunshine, J.~M., A'Hearn, M.~F., Groussin, O., et al.\ 2006, Science, 311, 1453


\bibitem[Tancredi et al.(2006)]{2006Icar..182..527T} Tancredi, G., Fern{\'a}ndez, J.~A., Rickman, H., \& Licandro, J.\ 2006, Icarus, 182, 527

\bibitem[Tancredi et al.(1994)]{1994A&A...286..659T} Tancredi, G., Rickman, H., \& Greenberg, J.~M.\ 1994, A\&A, 286, 659

\bibitem[Th{\'e}bault \& Beust(2001)]{2001A&A...376..621T} Th{\'e}bault, P., \& Beust, H.\ 2001, A\&A, 376, 621 

\bibitem[Thomas et al. (2015)]{} Thomas, N., Sierks, H., Barbieri, C., Lamy, P.L., et al.\ 2015, Science, 347, aaa0440


\bibitem[Urakawa et al.(2011)]{2011Icar..215...17U} Urakawa, S., Okumura, S.-i., Nishiyama, K., et al.\ 2011, Icarus, 215, 17

\bibitem[Volkov \& Lukyanov(2008)]{2008SoSyR..42..209V} Volkov, A.~N., \& Lukyanov, G.~A.\ 2008, Solar System Research, 42, 209

\bibitem[Warell et al.(1999)]{1999A&AS..136..245W} Warell, J., Lagerkvist, C.-I., \& Lagerros, J.~S.~V.\ 1999, A\&As, 136, 245

\bibitem[Weiler et al.(2011)]{2011Icar..212..351W} Weiler, M., Rauer, H., \& Sterken, C.\ 2011, Icarus, 212, 351 

\bibitem[Welsh \& Montgomery(2013)]{2013PASP..125..759W} Welsh, B.~Y., \& Montgomery, S.\ 2013, PASP, 125, 759

\bibitem[Wyatt et al.(2007)]{2007ApJ...663..365W} Wyatt, M.~C., Smith, R., Su, K.~Y.~L., et al.\ 2007, ApJ, 663, 365 

\bibitem[Zuckerman et al.(2010)]{2010ApJ...722..725Z} Zuckerman, B., Melis, C., Klein, B., Koester, D., \& Jura, M.\ 2010, ApJ, 722, 725 

\bibitem[Zuckerman et al.(2003)]{2003ApJ...596..477Z} Zuckerman, B., Koester, D., Reid, I.~N., H$^¨u$nsch, M.\ 2003, ApJ, 596, 477 

\end{thebibliography}
\end{document}